%% file: paper.tex
\documentclass[aps,prd,showpacs,preprintnumbers]{revtex4}
\usepackage{amsmath,amssymb}
\usepackage{color}
\usepackage{graphicx}
\usepackage{subfigure}
\usepackage{physics}
\usepackage{multirow}
\input{macros.tex}
\begin{document}

\title{Nonperturbative potential for study of quarkonia in QGP}
\author{Dibyendu Bala and Saumen Datta}
\email{dibyendu.bala@tifr.res.in,saumen@theory.tifr.res.in}
\affiliation{Department of Theoretical Physics, Tata Institute of Fundamental
Research, Homi Bhabha Road, Mumbai 400005, India.}

\begin{abstract}
  A thermal potential can be defined to facilitate understanding the
  behavior of quarkonia in quark-gluon plasma. A nonperturbative evaluation
  of this potential from lattice QCD is difficult, as it involves real-time
  corelation function, and has often involved the use of Bayesian analysis,
  with its associated systematics. In this work we show that using the
  properties of the static quarkonia thermal correlation
  functions, one can directly extract a thermal potential for quarkonia from
  Euclidean Wilson loop data. This leads to a controlled extraction, and
  allows us to judge the suitability of various model potentials. We also
  discuss the phenomenology of quarkonia in the gluonic plasma.
\end{abstract}
\pacs{11.15.Ha, 12.38.Gc, 12.38.mh, 25.75.Nq}
\preprint{TIFR/TH/19-32}
\maketitle

\section{INTRODUCTION}
\label{sec.intro}
Quarkonia, mesonic bound states of heavy quark and antiquark, have played 
a very important role in our understanding of the physics of strong
interactions. The experimental signatures of some of these states are
distinctive, the most iconic being the dileption peak of the vector
quarkonia. In the theoretical side, the 
heavy quark mass, $M_Q \gg \lqcd$, leads to simplifications. The
earliest insights about properties of quarkonia states were obtained by treating them as nonrelativistic states bound by a color electric potential. The 
potential suitable for studies of quarkonia has been calculated in detail using
numerical Monte Carlo studies on lattice-regularized QCD; see, e.g., Ref.
\cite{bali} for a review. The potential remains an important ingredient in 
a systematic expansion of quarkonia in $1/\mq$ \cite{pnrqcd}. 

The dilepton peaks of the vector quarkonia, in particular that of the $J/\psi$,
have been extremely important signatures of creation of quark-gluon plasma
(QGP) in ultrarelativistic heavy ion collisions (URHIC), following the
suggestion three decades ago \cite{satz} that 
the screening of the color charge inside QGP will lead to dissolution of bound 
states. This was made more quantitative in follow-up studies \cite{digal}. The 
early studies used a perturbative Debye-screened form,
\beq
\vre = - \frac{\alpha(T)}{r} e^{- \md r}
\eeq{debye}
which is the free energy of a static $Q \bar{Q}$ pair in perturbative QGP.
Here $\md$ is the Debye mass, $\alpha = \frac{\textstyle 4}{\textstyle 3}
\; \frac{\textstyle g^2}{\textstyle 4 \pi}$, and $g$ is evaluated at
a scale determined by
the temperature $T$. Nonperturbatively, the free energy of $Q-\bar{Q}$ 
pair in plasma was calculated using lattice QCD \cite{zantow}, which was
used as a proxy for
an effective finite temperature potential. However, in the early days a
proper formalism for potential-based study of quarkonia in QGP was missing.
In particular, other thermodynamic quantities can be derived from the free
energy, e.g., an ``internal energy'' for the $\qqb$ pair \cite{internal}; the
use of such quantities have also been explored in the literature \cite{wong}.

A theoretical formalism for an ``effective finite temperature potential'', 
that can be used to study experimentally observed quantities like the dilepton
rate, was first provided in ref. \cite{impot}. The starting point is a
point-split version of the dilepton current,
\beq
J^\mu_{\vec{r}}(t,\vec{x}) =
\bar{Q}\left(t,\vec{x}+\frac{\vec{r}}{2}\right) \gamma^\mu \; 
\wil\left(t; \vec{x}+\frac{\vec{r}}{2},
\vec{x}-\frac{\vec{r}}{2} \right) Q
\left(t,\vec{x}-\frac{\vec{r}}{2}\right), 
\eeq{pointsplit}
where $\wil$ is a suitable gauge connection such that $V^\mu$ is
gauge invariant, and the angular brackets denote thermal average.
Defining the correlation function
\beq
\cgt = \int d^3x
\; \left\langle J^\mu_{\vec{r}}(t,\vec{x}) \,
J_{\mu,\vec{r}}(0,\vec{x}) \right\rangle
\eeq{cgt}
the spectral function $\rjwr$ is defined from its Fourier transform, 
\beq
\rjwr \ = \ \left(1 \, - \, e^{-\omega/T}\right) \ \int dt \; e^{i \om t} \; \cgt.
\eeq{spectral}
The dilepton rate is proportional to the spectral function of the point
current, $\rw = \lim_{\vec{r} \to 0} \rjwr$.

Since we are interested in heavy quarks with $\mq \gg T, \, \lqcd$, \eqn{cgt}
simplifies. Going to the nonrelativistic notation
$Q=\dbinom{\psi}{\chi}$ where $\psi, \ \chi$ are nonrelativistic
fields that annihilate a quark and create an antiquark, respectively,
and remembering that since $\mq \gg T$, the thermal states do not
include $Q$ fields, the leading ($\mq^0$) term in an $1/\mq$ expansion
gives
\beq
\cgt \equiv \int d^3x \, \left\langle \chi^\dagger\left(t,\vec{x}
-\frac{\vec{r}}{2}\right) \sigma_k \; 
\wil^\dagger \psi\left(t,\vec{x}+\frac{\vec{r}}{2}\right) 
\ \psi^\dagger\left(0,\vec{x}+\frac{\vec{r}}{2}\right) \sigma_k \, 
\wil \, \chi \left(0,\vec{x}-\frac{\vec{r}}{2}\right)
\right\rangle.
\eeq{nrc}
If one has a system where the sole interaction term is 
a potential $V(\vec{r})$ between the quark and the
antiquark, then it is easy to show that, to leading
order in $1/\mq$, $\cgt$ satisfies \cite{blaizot7}
\beq
\left( i \, \partial_t \, - \, \frac{\nabla_{\vec{r}}^2}{\mq} \right) \cgt
\ = \  V(\vec{r}) \; \cgt . 
\eeq{nr}
In our theory where the $\qqb$ are interacting with the thermal
medium, we can then define a potential by equating the left hand side
of \eqn{nr} to $V(t, \vec{r}) \; \cgt$ (staying within leading order
of $1/\mq$), where the interaction effects
are summarized in a time-dependent $V(t, \vec{r})$. 
An effective thermal potential, $\vt$, can then be defined in the large $t$ 
limit, if the limit exists: $\vt = \lim_{t \to \infty} V(t,
\vec{r})$.

The potential $\vt$ can be 
obtained by going to the static limit, where, modulo renormalization factor, 
$\cgt$ reduces to a Minkowski-time Wilson loop:
\beq
W_M(t, \vec{r}) \ = \ \frac{1}{3} \, {\rm Tr} \;  \mathbb{P}
\, e^{i \int_0^t dt_1 A_0(t_1,\vec{r}/2)} \; \wil\left(t; \vec{r}/2,
  -\vec{r}/2\right) \; \mathbb{P} \, e^{i \int_t^0 dt_2 A_0(t_2, -\vec{r}/2)}
  \wil\left(0; -\vec{r}/2,\vec{r}/2 \right)
\eeq{wm}
and \eqn{nr} reduces to
\beq
i \, \partial_t \; \log W_M(t, \vec{r}) \xrightarrow[t \to \infty]{}
\vt,
\eeq{pot}
which defines our thermal potential \cite{impot,blaizot7}. Using $\vt$
to calculate $\cgt$ from \eqn{nr} will give the resummation of the
leading ladder diagrams.

A calculation of $\vt$ in leading order hard thermal loop (HTL)
perturbation theory gives \cite{impot}
\begin{eqnarray}
\vt &=& \vre \ - \ i \, \vim, \qquad {\rm where} \nonumber \\
\vim &=& \alpha \, T \times \int\limits_0^\infty dz \, 
\frac{2 z}{\left(z^2+1\right)^2} \left[ 1 - \frac{\sin z x}{z x} \right] 
\label{htl}
\end{eqnarray}
and $\vre$ is given in \eqn{debye}. In \eqn{htl} we have absorved a
negative sign in the definition of $\vim$, so that $\vim$ takes positive
values. $\vre$ corresponds to the usual
physics of Debye screening in medium, such that for sufficiently large
screening, the bound states will not form. On the other hand, $\vim$
clearly leads to a broadening of the spectral function peak. It
captures the physics of collision with the thermal particles leading
to a decoherence of the $\qqb$ wavefunction \cite{ar,akamatsu}. For
the quark and antiquark far apart, $r \gg T$, $\vim$ reaches a finite
limit $\alpha T$ giving the damping rate of the
individual quarks \cite{blaizot7}.

It is well-known that the perturbative calculation, \eqn{htl}, is not 
suitable at temperatures $\lesssim$ a few times $\tc$, the deconfinement 
temperature. The aim of this paper is to make a nonperturbative calculation
of an effective thermal potential, using numerical lattice gauge theory 
techniques. Following the insight of Ref. \cite{impot}, various authors have 
tried calculating the thermal potential nonperturbatively. In the next section 
we will outline our strategy. More details, and some discussion on difference
from earlier studies, can be found in \scn{results}.

The potential description, \eqn{nr}, is of course an approximate
description of in-medium quarkonia. First, here the $\qqb$ pair is
treated as an external probe put in an equilibrium plasma.
Then (in the perturbative language) it accounts for a subclass of
diagrams. At zero temperature, the justification for this is
well-understood. At finite temperature, extra scales come into play,
making the picture more complicated. A systematic, effective field
theory based study of the interplay of these scales has been made in
Ref. \cite{pnrqcdT} in perturbation theory. In the hierarchy of scales
\[ M \gg \pi T \gg 1/r_B \gtrsim \md \sim gT \gg E_B \] one gets the potential
\eqn{htl}, where $r_B$ is the radius of the bound state and $E_B$ the
binding energy. For the temperatures of interest in heavy ion
collision experiments, this hierarchy of scales is hardly satisfied.
The effective field theory version, however, is perturbative and
therefore cannot be directly used for phenomenology.

Instead of going through the nonrelativistic potential route, one
could instead try to directly calculate the spectral function by
studying the Euclidean $\langle J_\mu J_\mu \rangle$ correlation
function and try to extract the spectral function from it. This has
been attempted for charmonia \cite{mem} and, using NRQCD, for
bottomonia \cite{memnr1,memnr2}. Unfortunately, the extraction of spectral
function from the Euclidean correlator is a notoriously difficult
problem, and the systematics are large (see \cite{review} for a
discussion, and \cite{mocsy} for early comparison of potential model
results with results of \cite{mem}). Therefore a nonperturbatively
determined potential
continues to be important for quarkonia phenomenology; see, e.g.,
\cite{pheno}. In recent years, there have also been attempts to come
out of the picture of external probe in equilibriated plasma, by
treating the quarkonia in plasma as an open quantum system
\cite{akamatsu,blaizot15,brambilla,miura}. The potential remains an
important structure in such frameworks \cite{blaizot15,miura}. 

The plan of the rest of the paper is as follows. After explaining the 
calculational methodology in the next section, in \scn{lat} we will give 
the calculational details. \scn{results} will give our 
results for the potential. Some phenomenological discussions and implications 
of the potential obtained will be discussed in \scn{pheno}, and 
the last section will have a summary and discussion.

\section{Nonperturbative study of finite temperature potential}
\label{sec.def}

The potential $\vt$ is directly related to the Minkowski space
Wilson loop, \eqn{pot}. But in numerical Monte Carlo studies we work in
Euclidean space. At zero temperature, it is straightforward to
calculate the $\qqb$ potential from the Euclidean Wilson loop: 
\beq
W(\tau, \vec{r}) \xrightarrow[\tau \to \infty]{} C(\vec{r}) \,
e^{-\tau \, V(\vec{r})} .
\eeq{0Tform}

At finite temperature, the simple spectral decomposition outlined in
\eqn{0Tform} does not work. The first attempt to extract the $\qqb$
potential from $\wrt$ was carried out in Ref. \cite{rhs}. The 
spectral decomposition of the Minkowski-time loop leads to \cite{rhs}
\beq
\wrt \; = {\int_{- \infty}^\infty  d\omega
  \ e^{-\omega \, \tau} \ \rwr} \Rrightarrow \vt \; = \; - \partial_\tau \, \log \, \wrt \; = \;  \frac{\int_{- \infty}^\infty   d\omega
  \ \omega \ e^{-\omega \, \tau} \ \rwr}{\int_{- \infty}^\infty  d\omega
  \ e^{-\omega \, \tau} \ \rwr} .
\eeq{poteucl}
Bayesian techniques were used to extract $\rwr$ from $\wrt$, and
then calculate the potential using \eqn{poteucl}.

The reconstruction of $\rwr$ from $\wrt$ is a notoriously unstable
problem. To make matters worse, the quality of the Wilson loop data
deteriorates quickly at large $\tau$ (this problem can be
somewhat alleviated with recent numerical techniques \cite{multilevel}).
While very impressive technological improvements have 
occurred in the Bayesian analysis techniques, the results
obtained for potential still have stability issues or have large
errorbars, especially for $\vim$. The first calculations
\cite{rhs} employed a Bayesian analysis method similar to Maximum entropy
and fitted the spectral function peak with a Lorentzian form. The results
obtained, however, are substantially different from a later analysis \cite{bkr}
which is of similar philosophy but employs a slightly different Bayesian
analysis, and fits to a skew-Lorentzian form \cite{br1}. 
The state-of-the-art for calculations in the gluonic plasma follow a similar
methodology and can be seen in Ref. \cite{br2}. 
Studies have also been carried out for full QGP (i.e. with thermal quarks), 
both with a Lorentzian form of the spectral function \cite{pw} and 
using Bayesian reconstruction methods \cite{prw}. While the improvement 
in the analysis method has been impressive, the results still suffer from 
stability issues; in particular, it is not easy to disentangle the effects of
$\vim$ and $\vre$ in $\wrt$. 

In this paper we take a different approach. Let us motivate it by writing
\beq
\wrt \ = \ e^{w(\tau, \vec{r})} \ W_{\scriptscriptstyle T}(\beta/2).
\eeq{logdef}
The physics of $\vre$ is very similar to that of the zero temperature
potential, \eqn{0Tform}. We therefore expect the real part of the potential
to come from the part of $w(\tau, \vec{r})$ which has a linear behavior
around $\beta/2$: $\tilde{w}(\tau, \vec{r}) \sim - (\tau - \beta/2) \vre
+ ...$. We isolate the $\tilde{w}$ part by splitting $\wrt$ as follows:
\bea
\wrt &=& \wap \times \wpr, \nonumber \\
\wap &=& \sqrt{\frac{\wrt}{\wrb}}, \label{wsplit} \\
\wpr &=& \sqrt{\wrt \times \wrb}. \nonumber
\eea
We find that $\btau = \log \wap$ has exactly the behavior we were
expecting: $\btau \sim \left( \frac{\textstyle \beta}{\textstyle 2}
\, - \, \tau \right) \vre$ over a large range of $\tau$ around $\beta/2$.
We illustrate this in \fgn{plateau}, where $\btau/(\beta/2-\tau)$ is
plotted. We also checked that for configurations below $\tc$, where we
can extract the potential from the full wilson loop, $\wap$ gives the
same result but reaches the plateau sooner.

In order to understand the behavior of $\wpr$, we write a spectral
decomposition for $\atau \ = \log \wpr$:
\beq
\atau \ = \ \int_{- \infty}^\infty d\om \ \sw \ \frac{1}{2} \left( e^{-\om \tau} \,
+ \, e^{-\om (\beta - \tau)} \right) \ \ + \, {\rm \tau-independent \ terms}.
\eeq{argsplit}
To go to the potential, we follow the usual route of going to real time
$\tau \to i t$: 
\beq
i \partial_t A(i t) \ = \ 
\int_{- \infty}^\infty d\om \ \sw \ \frac{\om}{2} \left( e^{-i \om t} \, - \,
e^{-\om \beta} e^{i \om t} \right) .
\eeq{arg2}
The potential is obtained in the large time limit of \eqn{arg2}, when
the oscillating factors $\exp(\pm i \om t)$ ensure that only the $\om \to 0$
contribution to the integral survives. In this limit $\exp(\beta \om) \to 1$
and it is obvious from \eqn{arg2} that $A(i t)$ leads to an imaginary
potential. One can then extract the real and imaginary parts
of the potential from $\wap$ and $\wpr$ respectively \cite{impot}, 
\bea
\vre &=& \lim_{t \to \infty} i \, \partial_t \, \log \wap|_{\tau \to it} \nonumber \\
-i \, \vim &=& \lim_{t \to \infty} i \, \partial_t \, \log \wpr|_{\tau \to it}.
\label{vsplit}
\eea

The argument above is motivated by perturbative studies of the
potential, where the split \eqn{vsplit} has been noted \cite{impot}. 
Even with \eqn{vsplit}, it is not obvious that the extraction of the
potential from the Euclidean correlation function is simple;
\eqn{vsplit} involves large Minkowski time, while the nonperturbative
data that can be obtained from the lattice is in Euclidean time $\tau
\in [0, \beta)$. Successful extraction of potential from \eqn{vsplit}
  is contingent upon the contribution from the ``potential modes''
  dominating the behavior of the correlation functions $\atau, \btau$.
  Fortunately, this is what was found in the
  behavior of the nonperturbative data. As we already discussed above
  and showed in \fgn{plateau}, over a large range of $\tau$,
  $\wap \sim \exp(-c \tau)$, leading to a straightforward extraction of
  $\vre$ from the slope of the exponent.  We actually
  obtained very similar plateau in all our lattices. See \scn{real}
  for more discussion.

\bef
\centerline{\includegraphics[width=7cm]{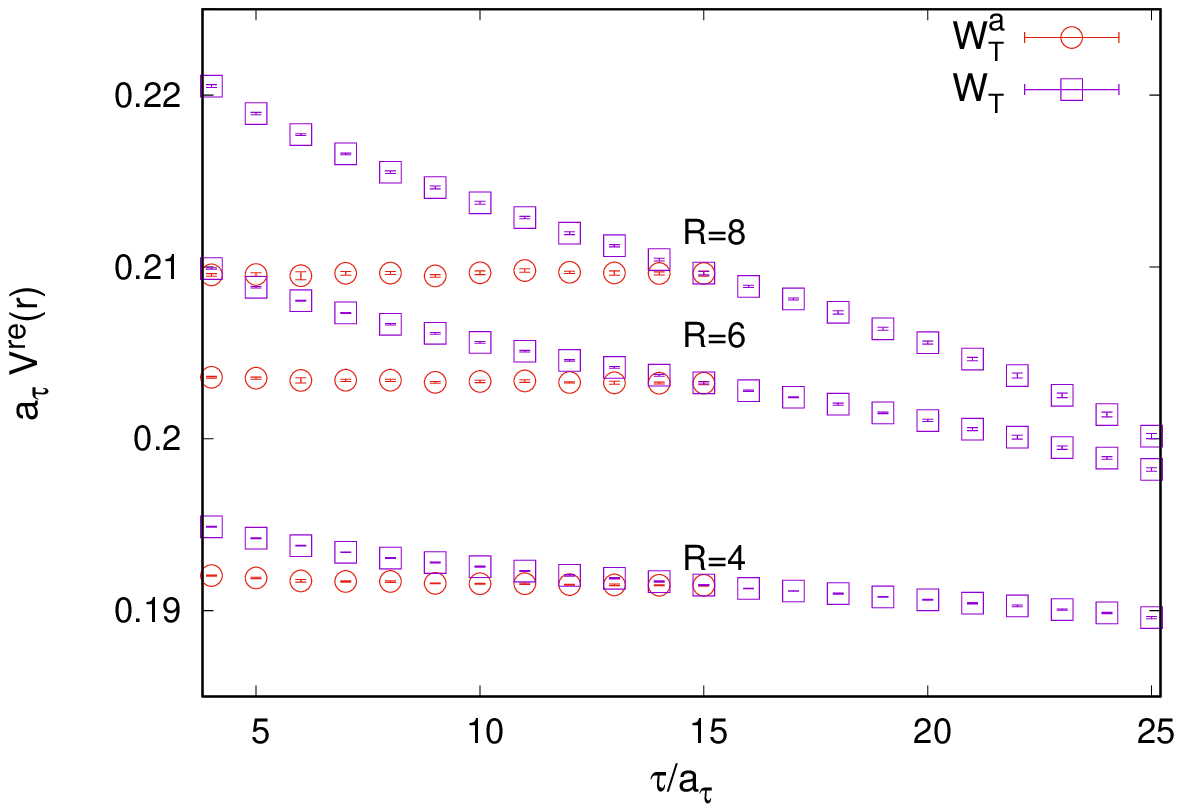}\hspace{1cm}
\includegraphics[width=7cm]{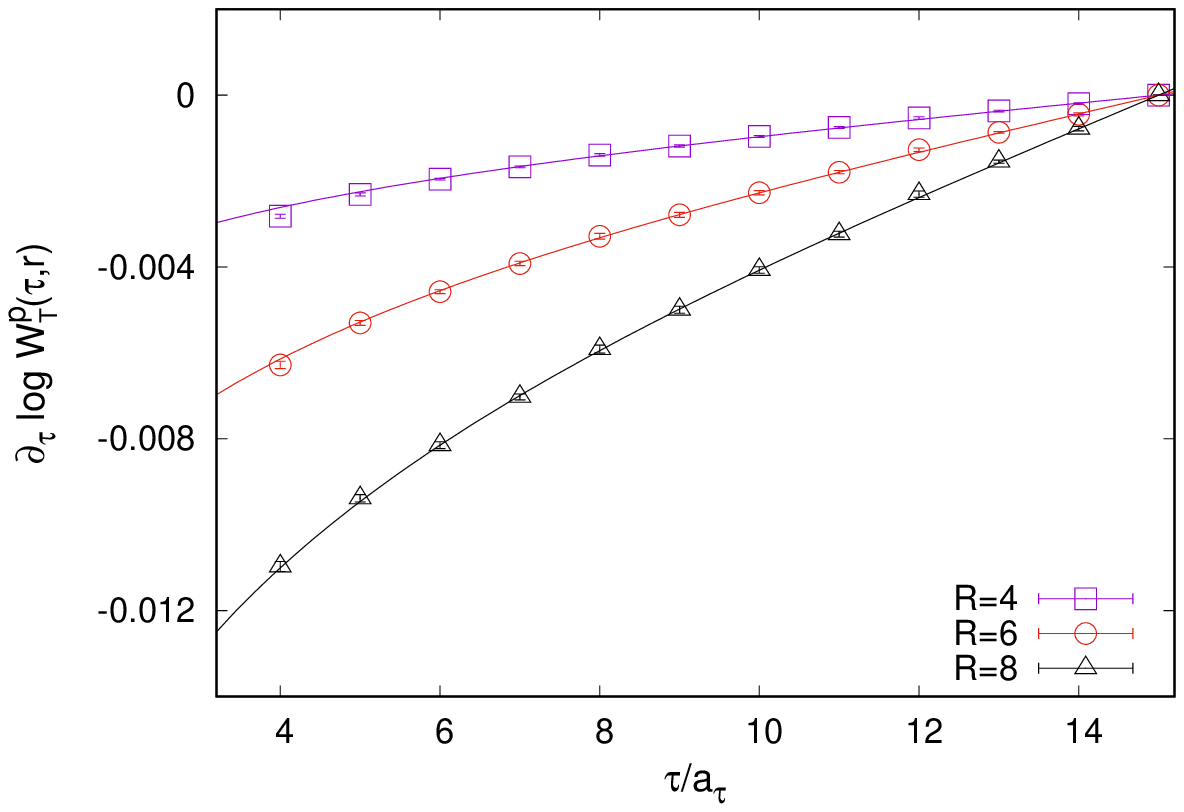}}
\caption{(Left) ``Local mass'' plot from $\wrt$ and $\wap$ for Set 3,
  1.5 $\tc$, at three different values of $R=r/a_s$. The results are from
  smeared Wilson loops.
(Right) $\partial_\tau \wpr$ for the same set; the lines show the singular
structure contribution (\eqn{imspectral}; see text).}
\eef{plateau}  

One, of course, does not expect such a simple behavior from $\atau$:
\eqn{argsplit} rules out a simple linear behavior near $\beta/2$.
This is expected: if $\wpr$ had a linear exponential
falloff, it would have contributed to a real potential! The large
time behavior of $W^p_{\scriptscriptstyle T}(\tau \to it, \vec{r})$
can be inferred from a closer examination
of \eqn{arg2}, using the fact that in the limit of large $t$, $\exp(-i
\, \om \, t) \, - \, \exp(i \, \om \, t \, - \, \om \, \beta)
\longrightarrow - 2 \pi \, i \, \omega \, \delta(\omega)$.
Then in order to get a potential
$-i \, \vim = \lim_{t \to \infty} i \, \partial_t \, A(it)$ we need 
\beq
\sw \ \underset{\omega \to 0}{\thicksim} \ \frac{1}{\omega^2} \;
\left( 1 + \mathcal{O}(\omega) \right).
\eeq{imspectral}
Interestingly, this leading singularity structure gives a very good
qualitative description of the $\tau$ dependence of $\partial_\tau
\atau$. This is illustrated in the right panel of \fgn{plateau}.

The argument in this section is based on the assumption that a
thermal potential can be defined using \eqn{pot}. We then make
plausibility arguments on the structure of $\wrt$, and show that the
nonperturbative lattice data supports this structure. The arguments
leading to \eqn{vsplit} can be made more concrete using Feynman
diagrammatic language \cite{blaizot7}: in \apx{pert} we outline this
argument. There we also show the results of the leading
order HTL perturbation calculation of $\wrt$ \cite{impot}, which
fully supports the structures of $\atau$ and $\btau$ discussed above,
and which motivated this nonperturbative study. The $1/\om^2$ behavior in
\eqn{imspectral} comes from the term $\frac{\textstyle
  \rho(\om)}{\textstyle \om^2}$ and a distribution function,
$(1 + \nbw) \xrightarrow{\omega \to 0} \frac{\textstyle T}{\textstyle
  \omega}$, which follows from the structure of the time-ordered
propagator (see \apx{pert} and \eqn{diag}). It is connected to the
scattering origin of the imaginary part of the potential, discussed
below \eqn{htl}.

Our strategy for extraction of the potential is therefore
straightforward: we extract $\vre$ from a linear fit to $\btau$ and 
to get $\vim$, we expand $\sw$ in \eqn{argsplit} in
the basis $(1+\nbw) \left\{1/\omega, \; \omega, ... \right\}$,
and extract $\vim$ from the coefficient of the most
singular term. As \fgn{plateau} suggests, the leading terms dominate
the data around $\beta/2$, allowing us to extract the potential
relatively simply. We discuss further details in \scn{results}. 

\section{Technical details of our study}
\label{sec.lat}
In this work, we have calculated the $\qqb$ potential in a gluonic
plasma, for moderately high temperatures $\le 2 \tc$. We have
generated lattices with a space-time anisotropic discretization with 
$\xi = \as / \at \approx 3$. A convenient algorithm for doing this 
is given in \cite{klassen}. We follow this reference to estimate the
lattice parameters we require. The anisotropy is estimated 
nonperturbatively from comparison of spatial
and temporal Wilson loops \cite{klassen}, while $\at$ is estimated from the
string tension calculated from temporal Wilson loops.
We use three sets of lattices, with
$\at$ ranging between $1/19 \tc$ and $1/45 \tc$. For each set, we
change the temperature by changing $\nt$, while keeping the
spatial volume fixed.

For each set, we first make short Monte Carlo runs at closely spaced
$\nt$ to find the $\nt$ for deconfinement transition. 
The final lattice sets used for the studies above $\tc$ are shown in
\tbn{sets}. For much of this
paper, we will measure all scales in units of $\tc$. However, for
\scn{pheno} we will need to quote physical units. We will do so by
taking the string tension $\sigma = (0.44 \; {\rm GeV})^2$. This
translates to a transition temperature $\sim$ 280 MeV.
The spatial extent of the lattices are 1.44 fm or above.
Some more details regarding the runs are given in \apx{sets}.

In order to determine the potential, we calculate thermal expectation
values of timelike Wilson loops, i.e., the Euclidean time version of
$W_M$ in \eqn{pot}. It is well-known that  for the spatial connections
$\wil$ straight thin-link gauge connections are not suitable: they  
lead to very noisy signals in numerical Monte Carlo studies. To
alleviate the problem due to extended spatial connections, we do
APE smearing \cite{ape}. This 
constitutes of a replacement of the elementary gauge links $U_i$,
\begin{eqnarray}
U_i (\vec{x}, \tau) &\rightarrow& {\rm Proj}_{SU(3)} \ \Bigl\{ 
\alpha \, U_i(\vec{x}, \tau) \ + \label{ape} \\ &{}& \sum_{\substack{1
    \le j \le 3 \\ j \ne i}} 
\bigl( U_j(\vec{x}, \tau) \,  U_i(\vec{x}+a_s \hat{j}, \tau)  \, 
U_j^\dagger(\vec{x}+a_s \hat{i}, \tau)
\ + \  U_j^\dagger(\vec{x}-a_s \hat{j}, \tau)  \, U_i(\vec{x}-a_s \hat{j}, \tau) 
\, U_j(\vec{x}-a_s \hat{j}+a_s \hat{i}, \tau) \bigr) \Bigr\} \nonumber
\end{eqnarray}
iteratively. The spatial connections $\wil$ are then constructed from
these smeared links. For this work, we have taken $\alpha$ = 2.5.

\bet
\setlength{\tabcolsep}{10pt}
\begin{tabular}{ccccccc}
  \hline
  Set & $\beta_s, \ \beta_t$ & $N_s$ & $N_t$ & $T/T_c$ & L(fm) &
  $a_t$(fm) \\
  \hline
\multirow{3}{*}{I} & \multirow{3}{*}{2.469, 14.8} &
\multirow{3}{*}{16} & 48 & 0.4 & \multirow{3}{*}{1.82} &
\multirow{3}{*}{0.038} \\
&         &      & 24 & 0.8 & & \\
&         &      & 16 & 1.2 & & \\
\\
\multirow{3}{*}{II} & \multirow{3}{*}{2.53, 15.95} & \multirow{3}{*}{24}
& 48 & 0.6 & \multirow{3}{*}{1.73} & \multirow{3}{*}{0.024} \\
   &      &       & 24 & 1.2 & & \\
&      &       & 20 & 1.5 & & \\
\\
\multirow{5}{*}{III} & \multirow{5}{*}{2.6, 16.98} &
\multirow{5}{*}{30} & 72 & 0.63 & \multirow{5}{*}{1.44} &
\multirow{5}{*}{0.016} \\
&      &       & 60 & 0.75 & & \\
&      &       & 38 & 1.2 & & \\
    &      &       & 30 & 1.5 & & \\
&      &       & 23 & 2 & & \\
\hline
\end{tabular} 
\caption{Parameter sets for the finite temperature runs.}
\eet{sets}

Note that \eqn{ape} does not involve the time direction, and the time
direction links are not smeared. So time slices and the definition of 
transfer matrix is not affected by the smearing. We use the multilevel 
algorithm \cite{multilevel} in the temporal direction: this allows us
to get a good signal even for Wilson loops with large
time extent. For calculation of the potential at $T=0$, smearing is routinely
used, and the potential should be independent of the smearing. In the
finite temperature case, the extracted ``potential'' may depend on the
details of the connection $\wil$; but the actual physical quantity one
is interested in, the quarkonia peak in dilepton channel, is
independent of it, as it is connected to the point current.
We do, however, do a detailed study of the
dependence of the potential on the smearing level
in the next section. 
 
In the literature, the correlator of Coulomb gauge fixed Wilson lines have 
often been used to extract the potential. The Coulomb gauge fixing can be 
formally understood as a dressing of the quark fields \cite{weise}:
\beq
\bar{\psi}(x) \; \psi(y) \vert_{\rm coul.} \equiv \bpom(x) \;
\pom (y)
\eeq{cgf}
where $\pom (x) = \Omega(x) \psi(x)$ and $\Omega(x)$ is a dressing 
function such that $\pom (x)$  is gauge invariant \cite{weise}. 

The Coulomb gauge potential has obvious advantages in that the
extended spatial links are not there.  At $T=0$, it is also easy to
argue (and has been well-tested) that the Coulomb gauge potential
agrees with the potential extracted from the Wilson loop. For $T >
\tc$ such detailed comparison does not exist in the literature. Here
we have made such a comparative study. The coulomb gauge is fixed to
an accuracy of $10^{-7}$. We have also checked that the results do not
change if the accuracy is made $10^{-6}$ or $10^{-9}$ instead.
The potential from this Wilson line correlator has also been
presented in \scn{results}. In particular for the imaginary part of
the potential, we observe differences between this potential and that
obtained from the smeared Wilson loop. Since the Wilson loop operator
does not involve dressing of the quark field, the connection to 
the point-point correlator at $\vec{r} \to 0$ is transparent.
We use the potential obtained from the Wilson loop for further studies
in \scn{pheno}.

\section{Potential calculated from Wilson loops}
\label{sec.results}
In this section we present the details of our extraction of the
potential, using \eqn{vsplit}. In \scn{real} we discuss the real part of
the potential. The results for the free energy of a $\qqb$ pair is
given in \scn{free}, and the extraction of $\vim$ is discussed in
\scn{imag}. Besides quoting the results for the potential, we also
compare the potential at different levels of smearing, and the results
for Coulomb gauge. Finally, in \scn{combo} we will discuss the
spectral representation \eqn{poteucl}, and touch on issues of
direct extraction of spectral function from Euclidean data.

\subsection{Real part of the potential}
\label{sec.real}
As outlined in \scn{def} and \fgn{plateau}, for smeared Wilson loops
the extraction of the real part of the potential from $\wap$ is
straightforward. Defining a local potential through
$- \partial_\tau \log \wap$ shows a plateau near
$\beta/2$. In the left panel of \fgn{reV-smearing} we show the
``local measurements'' of $\vre$ from Wilson loops at different levels
of smearing. The errorbars shown are from a Jackknife analysis, after
blocking the data to reduce autocorrelation. As the figure shows, 
while for a small number of APE smearing steps, the local mass takes
time to reach a plateau, on
increasing the number of steps a plateau is reached quickly, and 
we can easily extract the potential using a single exponent fit.
While we have shown the local mass for one particular case, the
effects are very similar for all our sets. 
For each smearing level the value obtained from the fit is shown by
the horizontal band of the same color. The goodness of the fit, as
demonstrated by $\chi^2$, is very good. The 
figure also shows that varying the number of smearing steps over a
large range does not seem to have any statistically significant effect
on the value reached at the plateau. 

\bef
\centerline{\includegraphics[width=8cm]{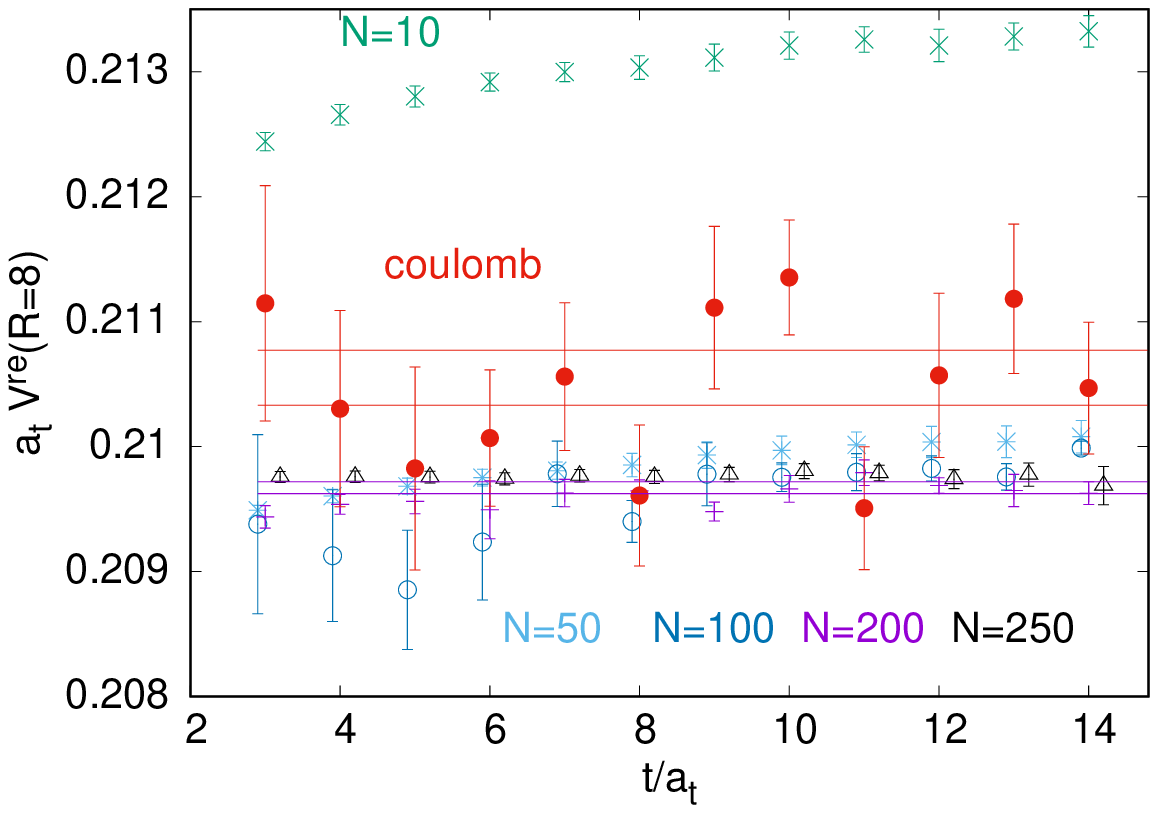}
  \includegraphics[width=8cm]{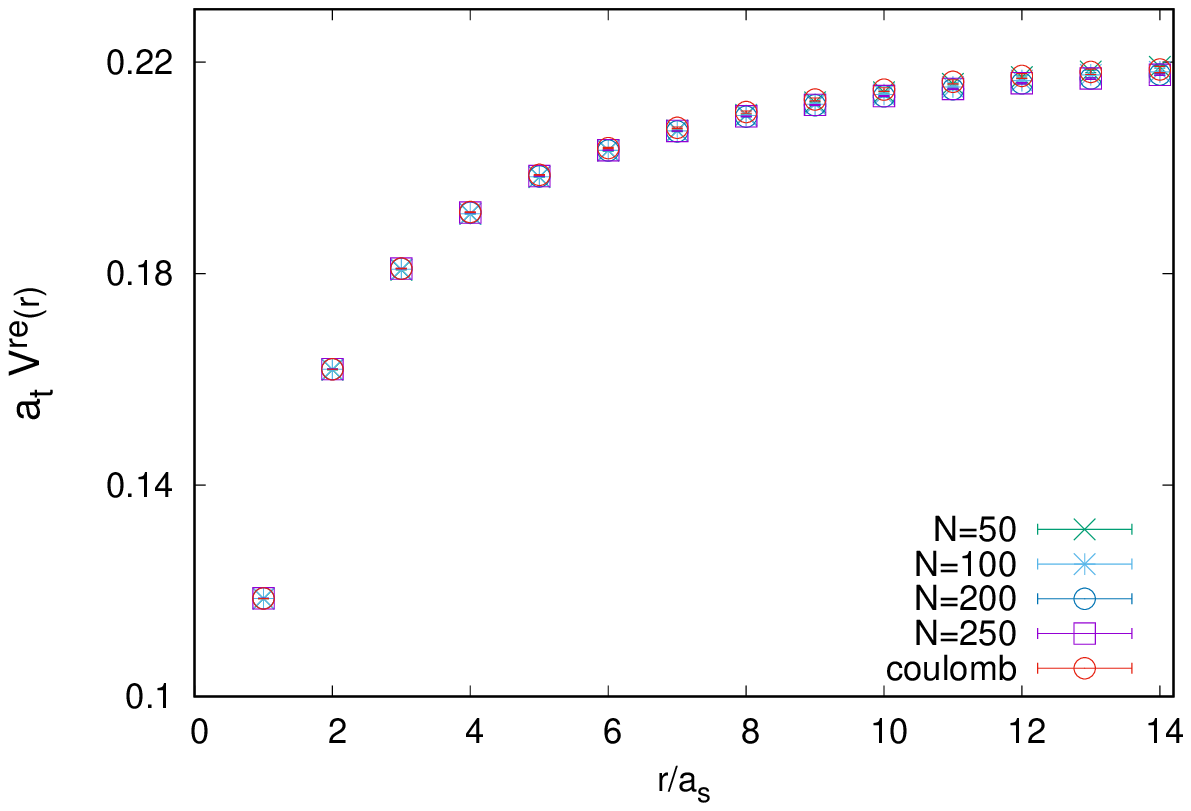}}
\caption{(Left) The ``local measurements'' of $\vre$ at $r=8 a_s$, at
  $T \sim 1.5 \tc$, for set 3. Shown are results from Wilson loops at
  different levels of smearing, and Coulomb gauge fixed Wilson lines.
  The horizontal bands show the result for the potential obtained from
  a single state fit. (Right) Estimates of $\vre$ for the same set,
  from Wilson loops at different levels of smearing, and for the
  Coulomb gauge fixed wilson line.}
\eef{reV-smearing}

We also show in the figure the local values of the potential obtained
from the Coulomb gauge fixed Wilson lines. As the figure shows, the
Coulomb gauge data seems to be noisier than the data from Wilson
loops. We checked that this is not an artifact of the accuracy at
which the Coulomb gauge is fixed. Also the Coulomb gauge results are
found to be close to the results from the smeared Wilson loops, but
the difference between them is statistically significant.

In the right panel of \fgn{reV-smearing} we summarize the fitted value
of $\vre$ for this set. At this scale, the dependence of the potential
on the smearing level is hardly visible. Similarly, the potential
from smeared Wilson loops and that from Coulomb gauge fixed Wilson
lines are very close, though they differ at $1 \sigma$ level.

As we have discussed in \scn{lat}, we believe that for study of
quarkonia property in medium, the potential from the smeared Wilson
loop is appropriate. It is satisfactory that $\vre$ becomes practically
independent of the level of smearing very soon. Anyway, when quoting a
result for $\vre$, we include, as a systematic error, some variation
with the level of smearing: for example, for the set shown in
\fgn{reV-smearing} we include the spread in results between smearing
levels of 100 and 250 as a systematic error. In what follows, our
error bars for $\vre$ include this variation for all sets.  

Results from lattices at a finite lattice spacing have discretization
errors. We can have an idea of the size of the discretization error by
comparing the results at different lattice spacings. As \tbn{sets}
shows, we have lattices with three different lattice spacings at 1.2
$\tc$, and at 1.5 $\tc$ we have results with two different lattice
spacings. In \fgn{cont-real} we show the
potentials calculated from lattices at different lattice
spacings. Within our error bars the results agree very well,
indicating that the cutoff effects are very small at these lattice
spacings. We will, therefore, take the results on our finest lattice
spacings as a valid estimator of the continuum results.
\bef
\centerline{\includegraphics[width=8cm]{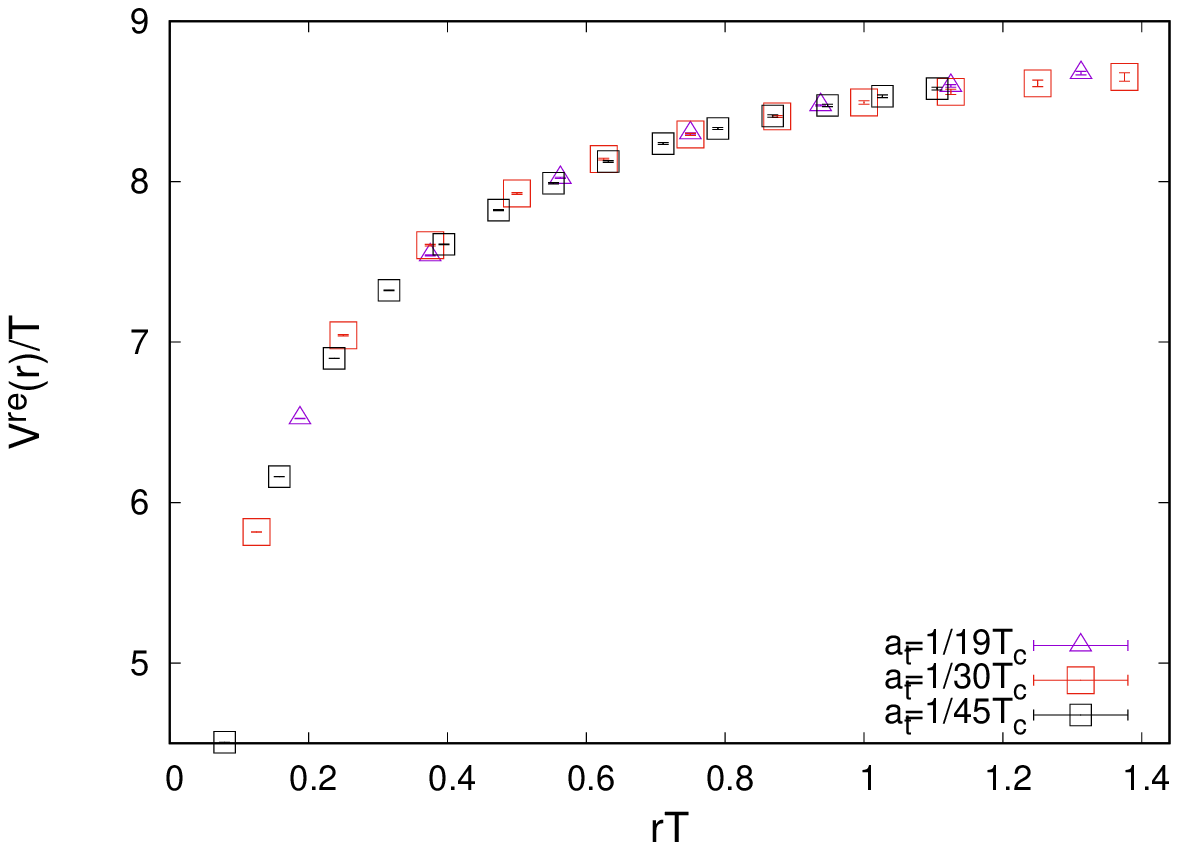}
  \includegraphics[width=8cm]{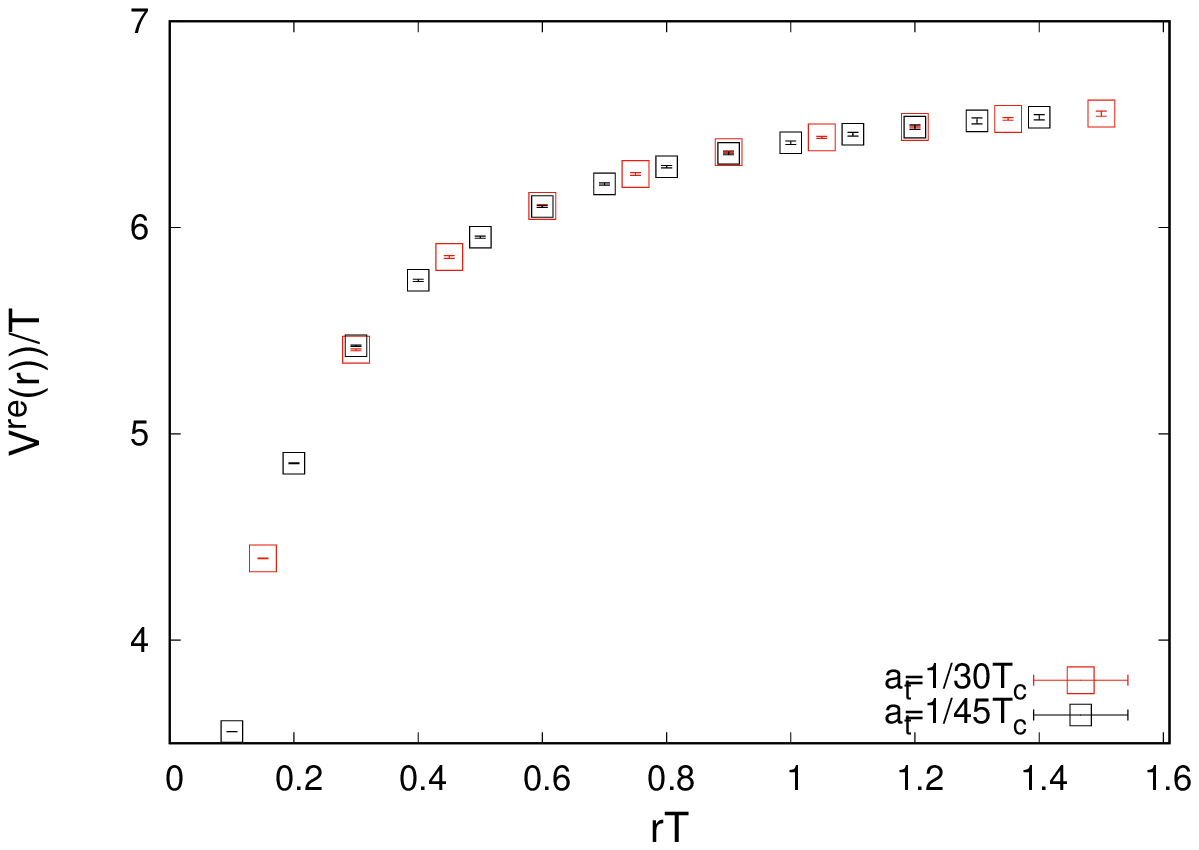}}
\caption{The potential $\vre$ at (left) 1.2 $\tc$, measured on
  lattices with three different lattice spacings; and (right) at 1.5 $\tc$,
  at two different lattice spacings.}
\eef{cont-real}

\fgn{pot-real} summarises our results for $\vre$ at different
temperatures. We see that the potentials at the two temperatures
below $\tc$ agree completely, indicating that the temperature effect
is small even at 0.75 $\tc$. The potentials have the familiar Cornell
form, with a dip at small $r$ and a linearly rising part for $r
\gtrsim$ 0.5 fm. This behavior changes abruptly on crossing $\tc$:
while the short distance part, $\lesssim$ 0.2 fm, remains similar to
the form below $\tc$, beyond $r \tc \sim 0.5 \sim$ 0.35 fm the effect
of string breaking clearly shows up, and the potential becomes flatter
with increasing temperature.

\bef
\centerline{\includegraphics[width=8cm]{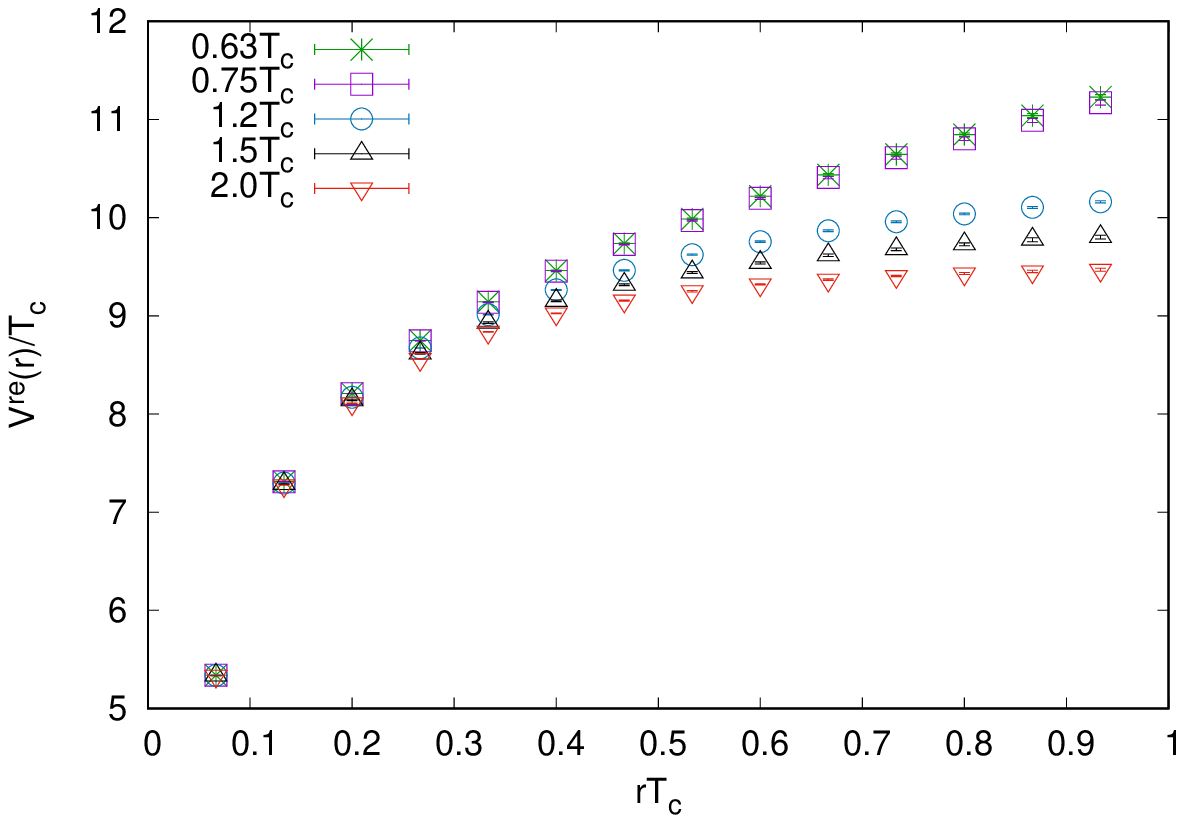}}
\caption{$\vre$ calculated from smeared Wilson loops. The results are
  from Set 3, which can be taken as a good estimate of the continuum
  results. \\
}
\eef{pot-real}

\subsection{Free energy}
\label{sec.free}
The study of the free energy cost of introducing a $\qqb$ pair in the
plasma is almost as old as the study of deconfinement transition in
QCD. The free energy of $\qqb$ pair was calculated from the correlator
of Polyakov loops, $\langle L(\vec{r}) \ L^\dag(\vec{0}) \rangle$
\cite{mclerran}. Later, the free energy cost of a singlet $\qqb$ pair
was connected to the cyclic Wilson loop (for sufficiently smeared
loops) \cite{circular}: 
\beq
\fre \ = \ -T \, \log W_{\scriptscriptstyle T}(\beta,
\vec{r})
\eeq{free}
or from Coulomb gauge fixed Circular Wilson lines \cite{nadkarni}
(see also \cite{owe}). In leading order perturbation theory,
the singlet free energy agrees with $\vre$.

The singlet free energy has been studied in great detail, for both
gluonic plasma and the theory with quarks \cite{zantow}, and we do not
intend to add to the existing results. Here we will, however, examine
the issue of whether the perturbative agreement between the free
energy and $\vre$ is also valid nonperturbatively.

\bef
\centerline{\includegraphics[width=8cm]{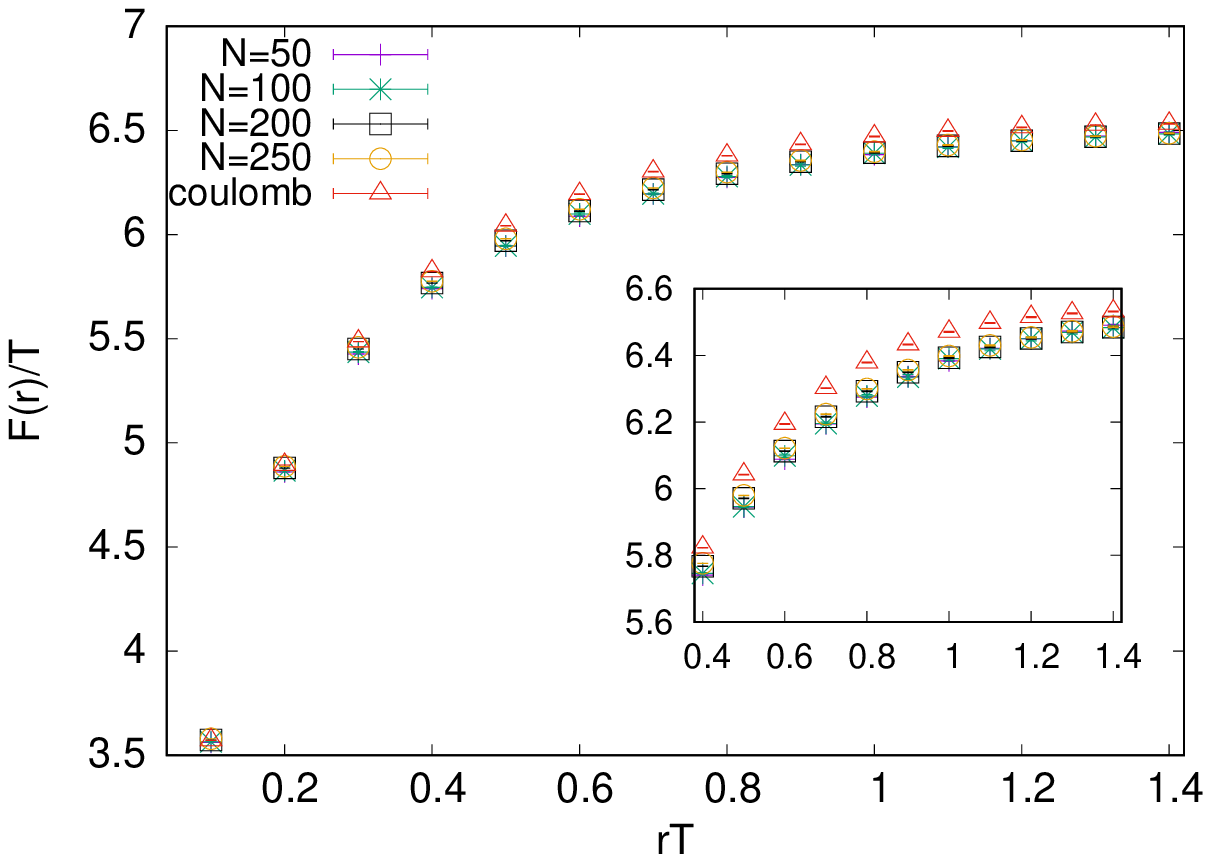}} 
\caption{Free energy calculated from cyclic Wilson loops at different levels
  of smearing, and from Coulomb gauge fixed circular Wilson lines, at 1.5 $\tc$
for set 2. The inset highlights the long distance part. The Coulomb
gauge result is seen to be close to that obtained from smeared Wilson
loop, but with statistically significant difference.}
\eef{free}

In \fgn{free} we show the singlet free energy
calculated from the smeared circular Wilson loop at different levels
of smearing, and that from the Coulomb gauge fixed operator. The
smearing dependence is similar to
what was seen for $\vre$: the results are quite insensitive to the
smearing level used. The Coulomb gauge operator is close to the Wilson
loop results, but not exactly identical.

In \fgn{free-Vre} we compare the free energy and $\vre$, extracted
from the smeared Wilson loops, at three different temperatures. As
discussed before in \scn{real}, the results are expected to
be valid continuum results. At all temperatures, we find that $\fre$
and $\vre$ are very close to each other. However, at long distances
$\vre$ shows slightly less screened behavior than $\fre$.

\bef
\centerline{\includegraphics[width=6cm]{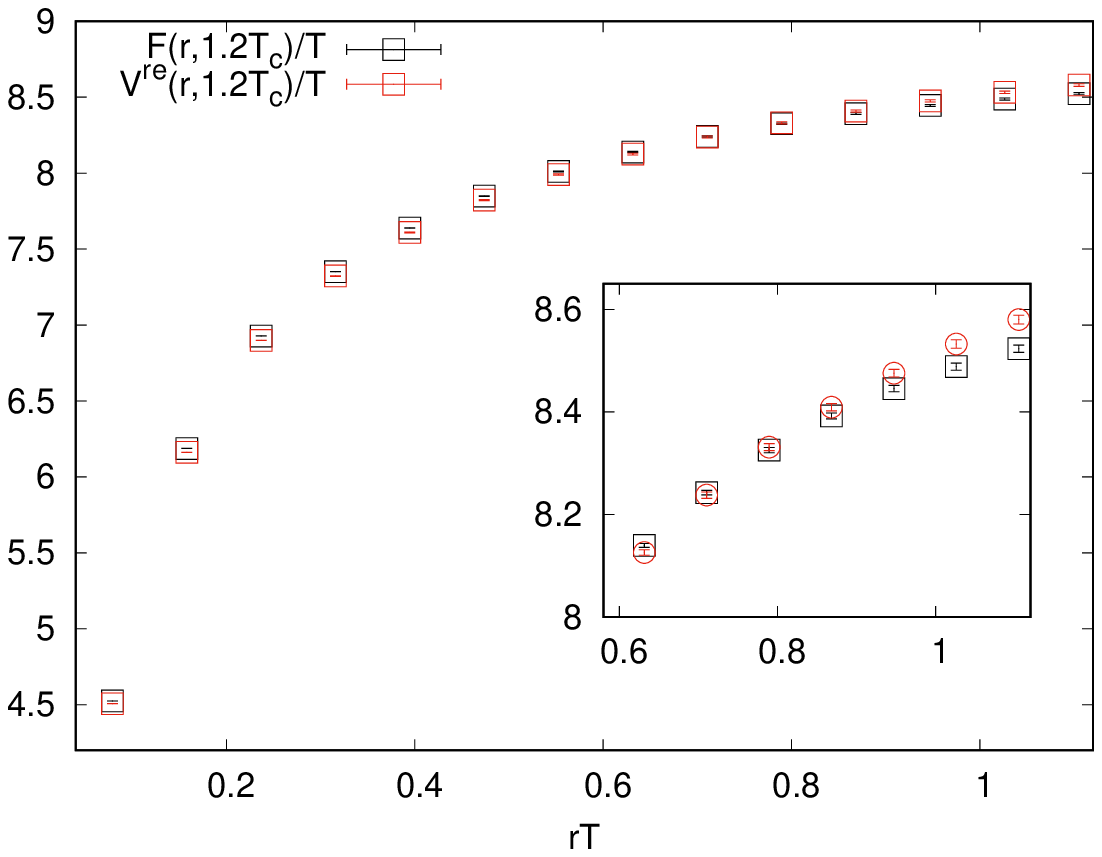}
  \includegraphics[width=6cm]{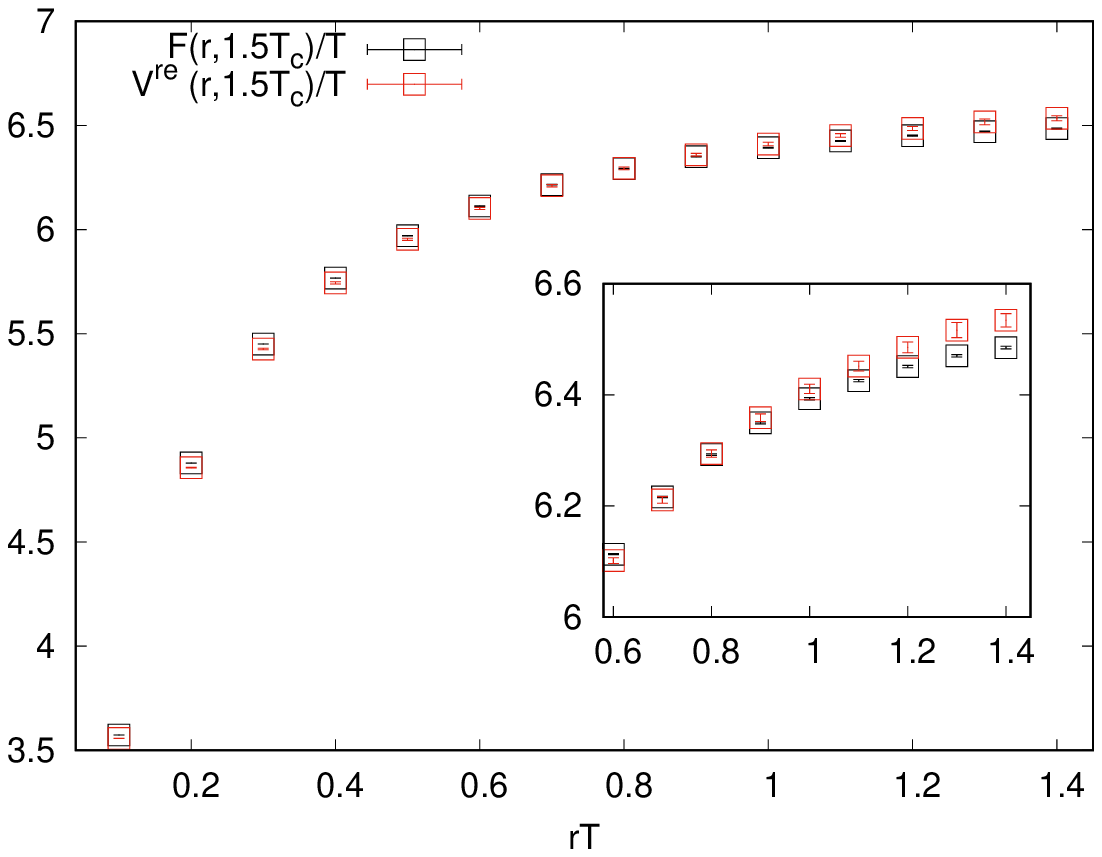}
  \includegraphics[width=6cm]{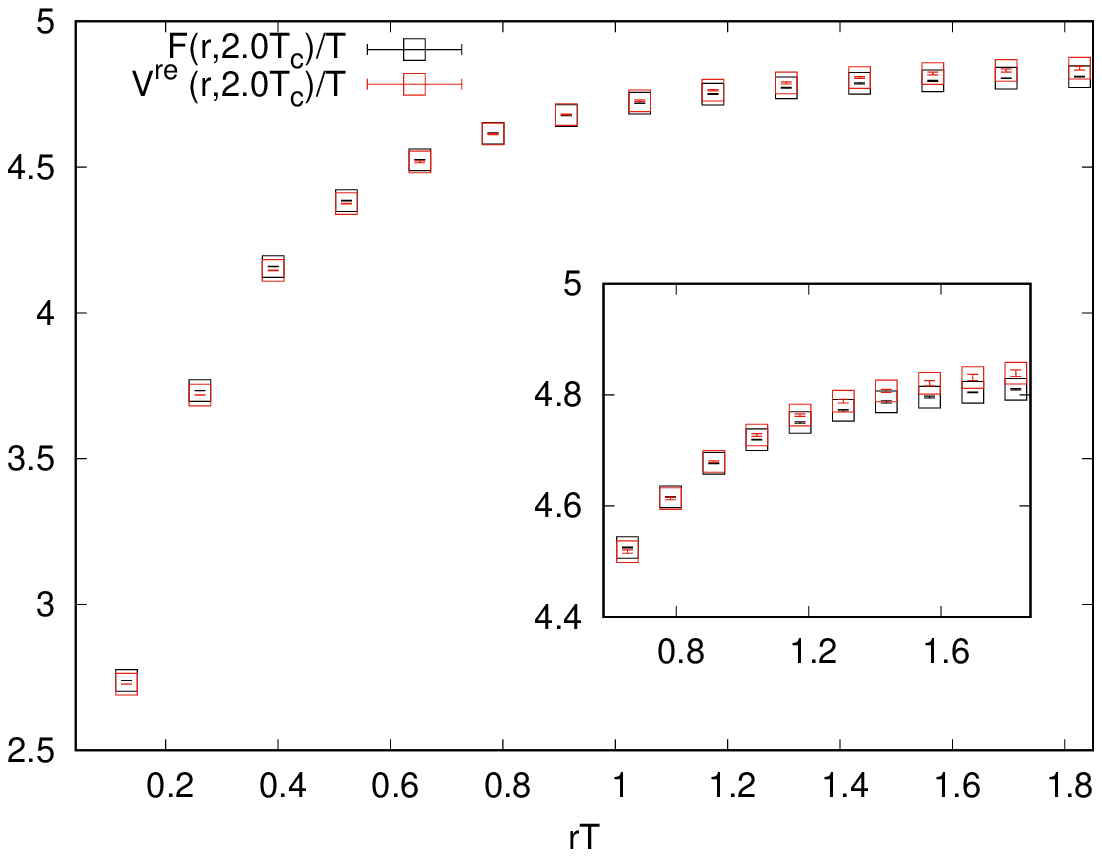}}
\caption{The free energy, \eqn{free}, 
  compared with $\vre$ at 1.2 $\tc$ (left), 1.5 $\tc$ (middle) and
  2 $\tc$ (right). The long distance part is highlighted in the inset.}
\eef{free-Vre}

\subsection{Imaginary part of the potential}
\label{sec.imag}
As we have discussed in \scn{def}, the behavior of the symmetrized
correlation function $\wpr$ is dominated by the most singular behavior
in \eqn{imspectral}, which is the term that corresponds to
$\vim$. Encouraged by this, we expand $\sw$ in \eqn{argsplit} in a
series
\beq
\sw \ = \ \left(1 \, + \, \nbw \right) \ \left( \frac{c_0}{\om} \
+ \ c_1 \, \om \ + \ c_2 \,
\om^3 \ + \ ... \right) .
\eeq{imseries}
where the form of \eqn{imseries} is motivated by the structure of
$\atau$ (see \eqn{diag} and the discussion at the end of \scn{def}). 
$\sw$ in \eqn{imseries} has the property that $\sigma(-\om;T)=
e^{\textstyle - \beta \om} \; \sw$ and so the integrand in \eqn{argsplit}
is an even function of $\om$; the even powers of $\om$ are absent
in \eqn{imseries} as they won't contribute to the integral.
The imaginary potential $\vim$ is obtained from the
coefficient of $1/\om$ term: $\vim =
\frac{\textstyle \pi}{\textstyle \beta} \, c_0$.
Putting \eqn{imseries} in \eqn{arg2}, we
get the linear series for the ``local mass'':
  \bea
  \partial_\tau \; \atau &=& c_0 \, \tilde{G}_0(\tau) \ + \ 
  \sum_{l=1,2,...} \, c_l \, \tilde{G}_l(\tau) \label{nbseries} \\
   \tilde{G}_0 &=& - \frac{\pi}{b} \; \cot \frac{\pi \tau}{\beta} \nonumber \\
  \tilde{G}_l &=& \frac{(2 l)!}{\beta^{2 l+1}} \left( \zeta \left( 2
  l+1, 1-\frac{t}{\beta} \right) \ -  \zeta \left( 2 l+1,
  \frac{t}{\beta} \right) \right) \nonumber
  \eea
where the generalized $\zeta$ functions $\zeta(s,x) = {\displaystyle
  \sum_{n=1}^\infty} \frac{\textstyle 1}{\textstyle (x+n)^s}$.
Note that this form \eqn{nbseries} is similar to, and could also be
motivated by, perturbation theory \cite{impot}.

The data near $\beta/2$ gives a very good fit to just two terms in
\eqn{nbseries}, and with three terms, almost the entire range of
$\tau$ could be fit in all our data sets.  In \fgn{im-smear} we show
the results for $\vim$ obtained with different levels of smearing. The
error bar here includes the variation due to change in number of terms
of \eqn{nbseries} in the fit.  The dependence on the level of smearing
is stronger here, but a plateau can be reached after some levels of
smearing. When quoting a result for the imaginary part of the
potential in what follows, our error bar encompasses the spread among
the different smearing levels in this plateau. 

\bef
\centerline{\includegraphics[width=7.5cm]{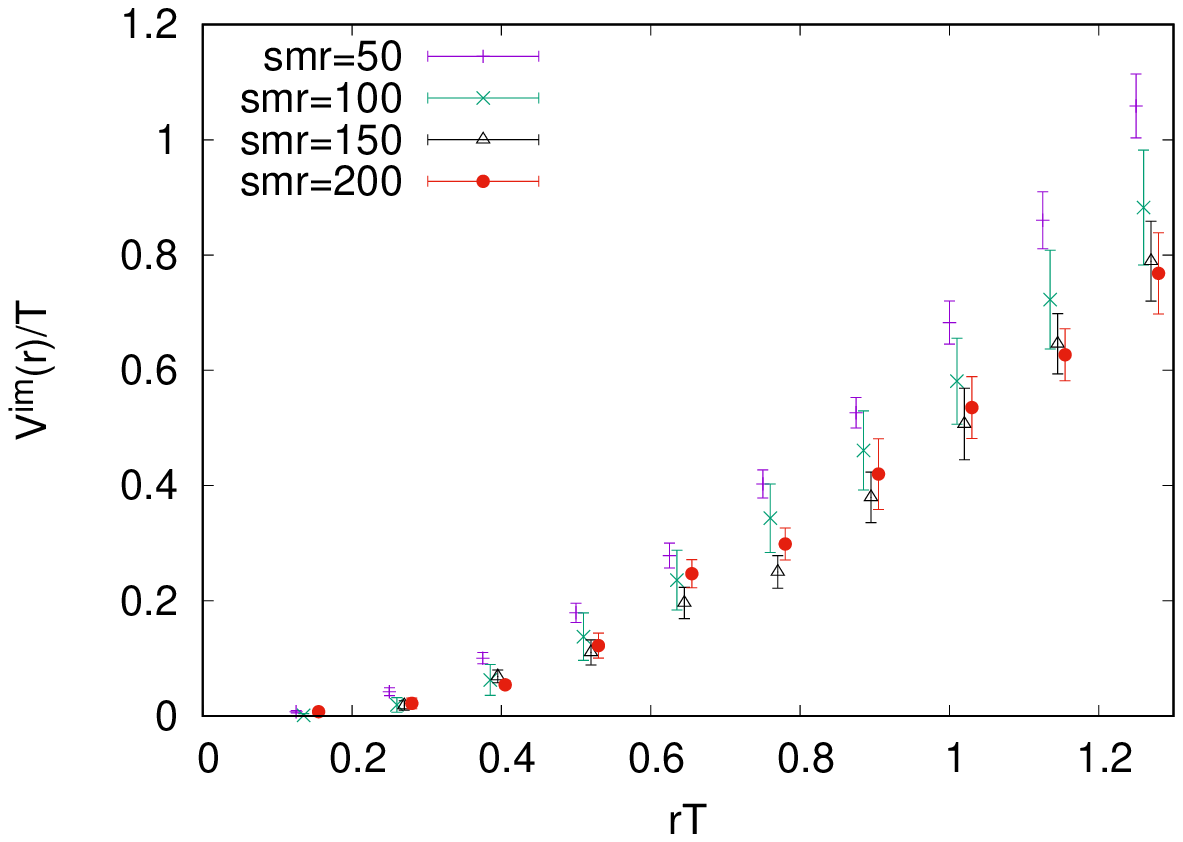}
 \includegraphics[width=7.5cm]{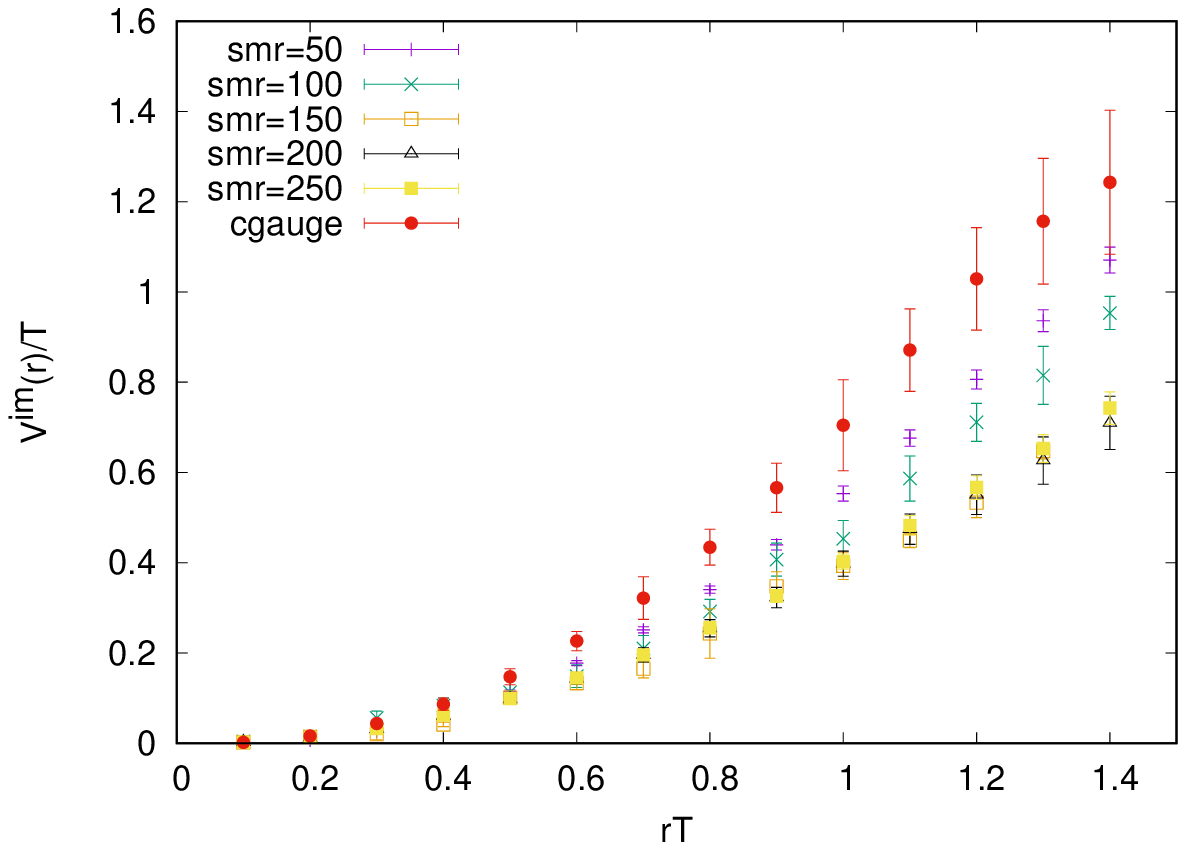}}
\caption{The imaginary part of the potential, $\vim$, for different
  smearing levels, (left) at 1.2 $\tc$, Set 2, and (right) at 1.5
  $\tc$, Set 3.} 
\eef{im-smear}

In \fgn{im-cont} we show the imaginary potential at two different
temperatures, obtained on lattices with different cutoffs. While our
coarsest lattice, set I, seems to show some lattice spacing
dependence, the results from the two finer sets agree very well. We
therefore take $\vim$ obtained from our finest lattice as a good
approximation to the continuum result.

\bef
\centerline{\includegraphics[width=7.5cm]{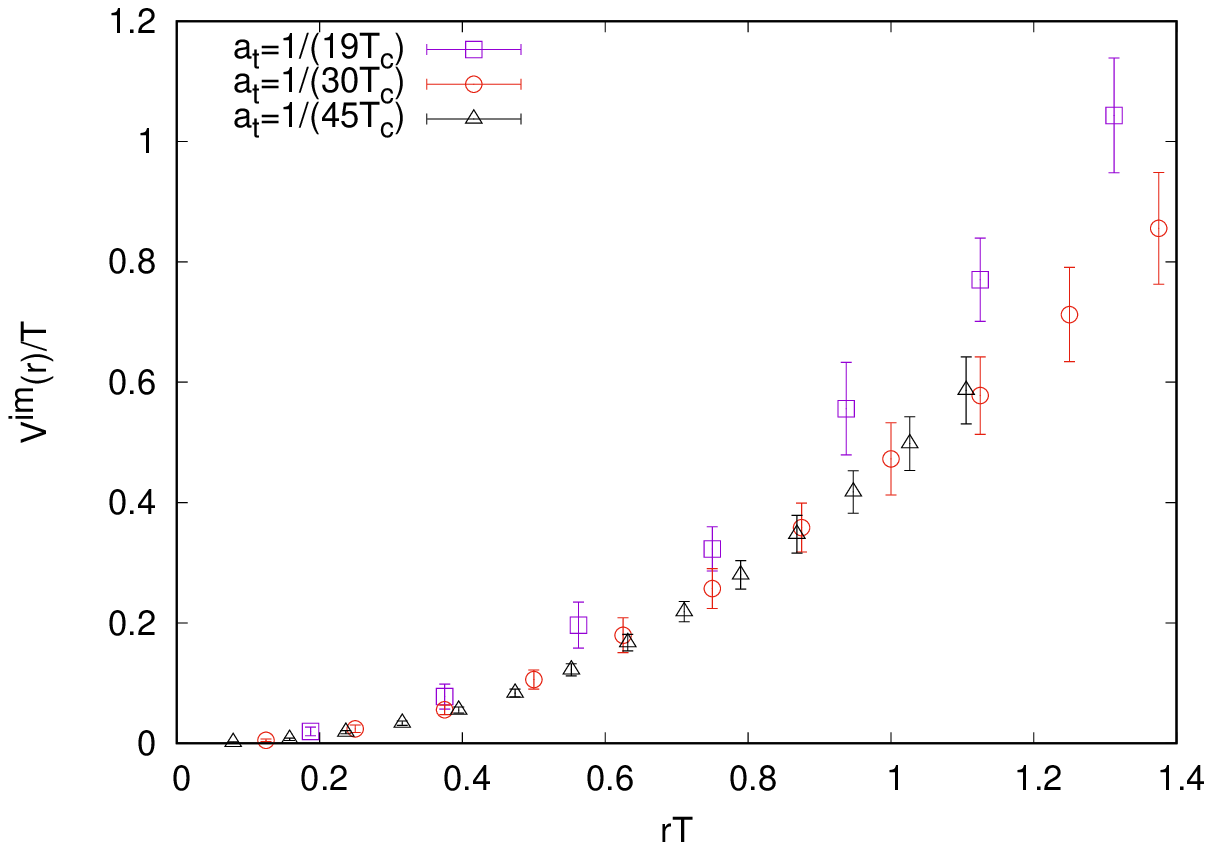}
 \includegraphics[width=7.5cm]{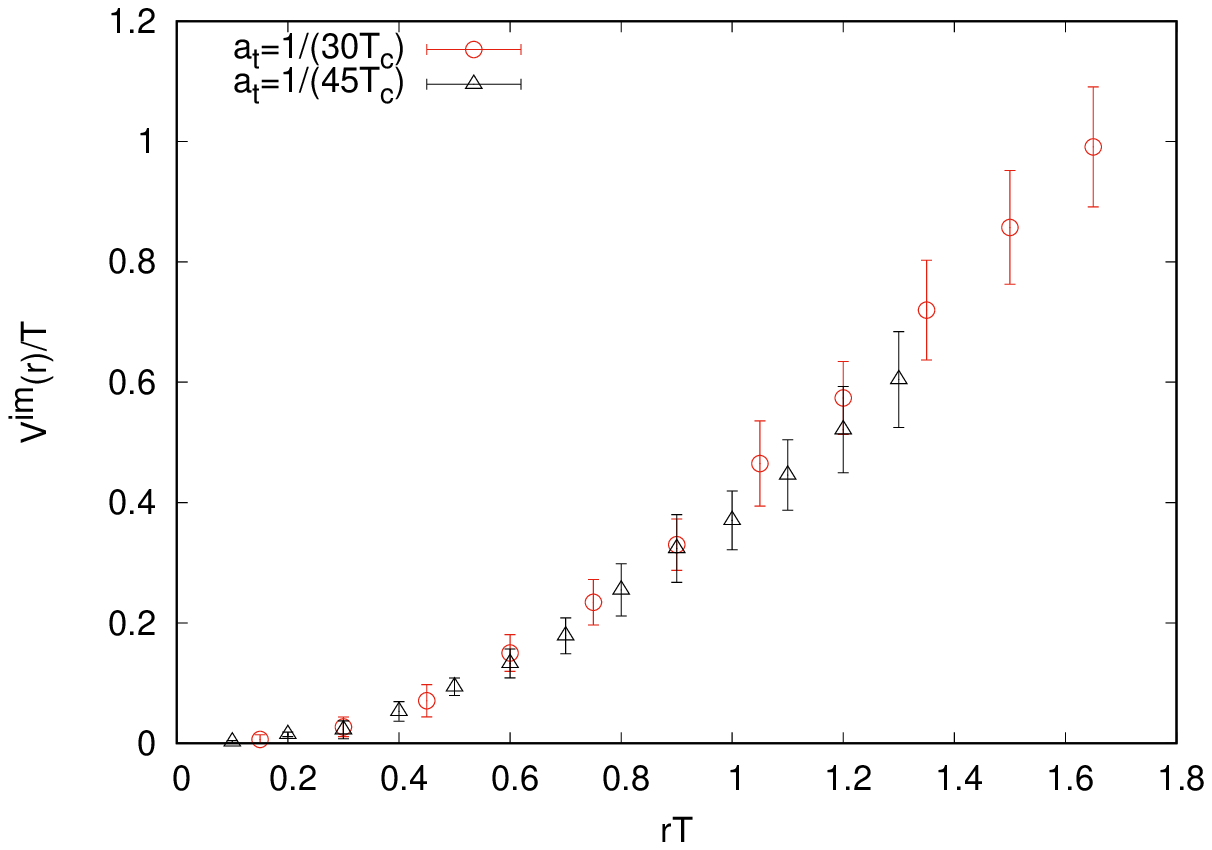}}
\caption{The imaginary part of the potential, $\vim$, at 1.2 $\tc$
  (left) and at 1.5 $\tc$ (right), at different lattice spacings. The
  results from different lattice spacings vary very little and so we take
  the results
  from our finest lattices as a good approximation to the continuum result.}
  \eef{im-cont}

  In \fgn{pot-im} we show our final results for the imaginary potential at
  three different temperatures. In \scn{pheno} we will use this data
  as the nonperturbatively evaluated $\vim$, and explore its
  physics. We have shown here the results above $\tc$ only; we have,
  however, run the same analysis strategy on the configurations below
  $\tc$, and checked that the results are consistent with zero, as expected.
  
\bef
\centerline{\includegraphics[width=8.5cm]{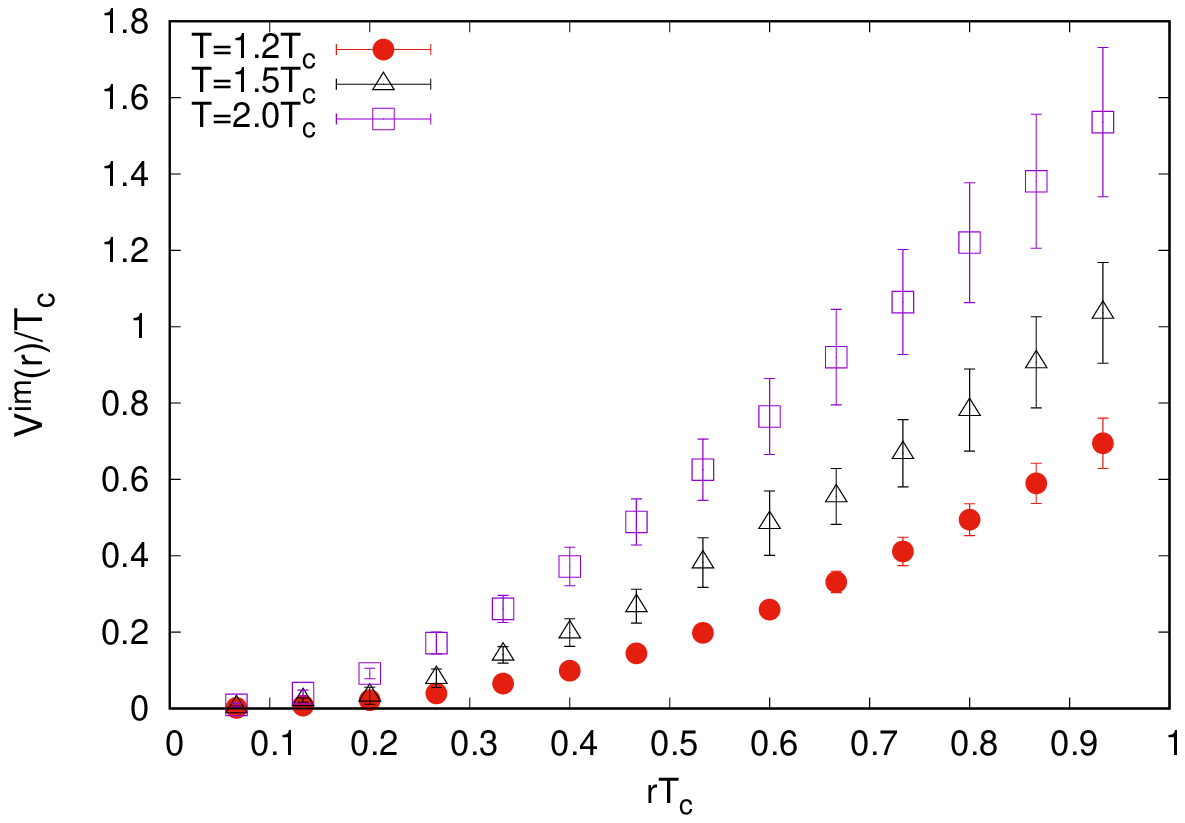}}
\caption{The imaginary part of the potential, $\vim$,
  extracted from set 3, at three different temperatures.}
\eef{pot-im}

\subsection{Low-$\om$ structure of the spectral function}
\label{sec.combo}
Combining the results of \scn{real} and \scn{imag}, we can
write the correlation function near the center of the lattice as
\beq
\wrt \ = \ e^{- \vre \left(\tau \, - \, \frac{\beta}{2}\right) \; - \;
  \frac{\beta}{\pi} \, \vim \, \log  \sin \left(\frac{\pi \, \tau}{\beta}
  \right) - ....} \ W_{\scriptscriptstyle T}(\beta/2, \vec{r})
\eeq{combo}
where the higher order terms,
\[ .... \ = \ \sum_l c_l \int_\frac{\beta}{2}^\tau \; \tilde{G_l}(\tau) \]
do not contribute to the potential. For explaining the Wilson loop data
over a substantial range near the center, just $c_1$ is enough, while
adding $c_2$ allows us to explain $\wrt$ over the entire range except a
couple of points at the edge.

Further insight into the potential can be obtained if we investigate the
structure of the low $\omega$ part of $\rwr$ in \eqn{poteucl}. In order to
do this, we take the Fourier transform of the structure of $\wrt$, \eqn{combo},
continued to real time:$W_{\scriptscriptstyle T}(t=-i \tau, \vec{r})$.
This shows a peak structure at low $\omega$, as has been anticipated in 
various lattice extractions of the potential, e.g., \cite{rhs,bkr,br1,br2,prw}.
Interestingly, however, the peak structure is very different from what has
been often anticipated. In the literature often a Lorentzian or a
Gaussian structure has been assumed for the peak. Instead, we find a
structure that is exponentially falling in the low $\om$ side of
the peak, $\sim \exp(\om/T)$, while in the high $\om$ side it falls only
like a power law. Illustration of the peak structure is shown for a few
representative values of $r$ in \fgn{spectralpeak}. Given this peak structure,
we could rephrase our discussion of the potential extraction by simply
starting from a structure like those shown in \fgn{spectralpeak}, and
extracting the potential from them. We checked numerically that the laplace
transform of the peak gives a statistically satisfactory description of
$\wrt$ near $\beta/2$. While the direct Bayesian inversions have to grapple
with the issue of convergence of the integral in the negative $\omega$ side,
here we could easily do the integral by putting a lower cutoff: because of
the sharp fall, the effect of the cutoff on the value of the integral is
negligible. The addition of the correction terms do not
have any significant effect on the position or the half-width of the peak,
but modifies the fall-off with $\om$ away from the position of the peak.

\bef
\centerline{\includegraphics[width=7.5cm]{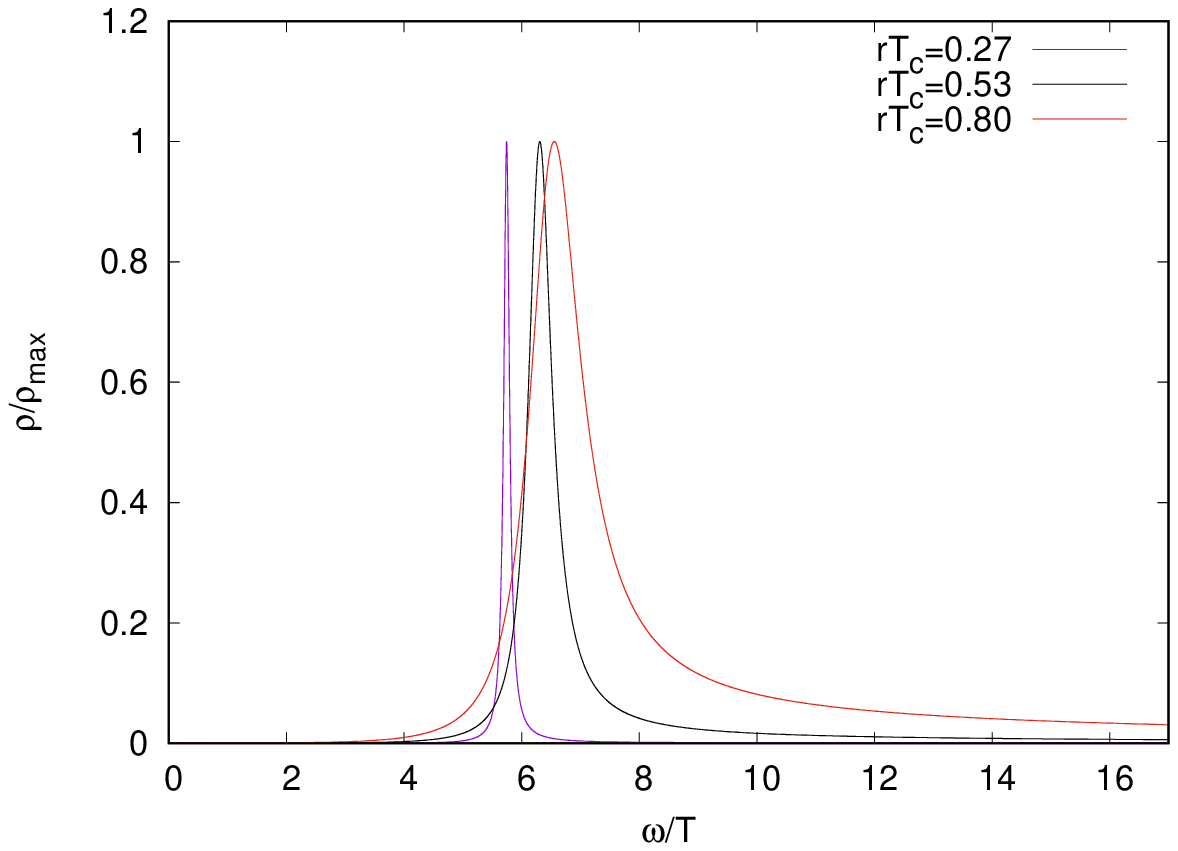}
  \includegraphics[width=7.5cm]{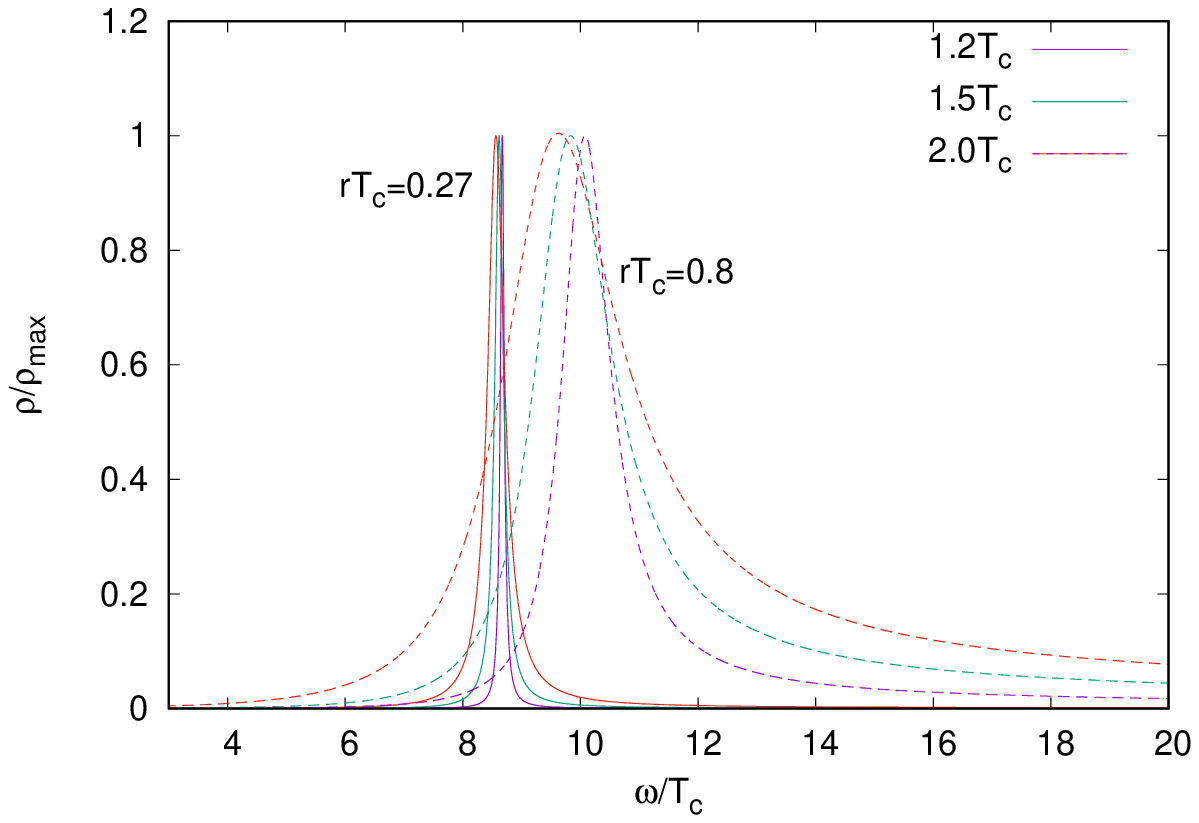}}
\caption{(Left) The low $\omega$ structure of the spectral function
  obtained from \eqn{combo}, at 1.5 $\tc$. Results for three
  representative values of $r$ are shown. (Right) The temperature
  dependence of the low $\omega$ peak. results for two values of $r$
  are shown.}
\eef{spectralpeak}

Bayesian statistics based studies of the potential proceed without
making strong assumptions about the structure of the peak. In fact,
some of the Baysian analyses use only very mild information about the
peak. We would like to add a note of caution here. If we do not make
the assumption \eqn{nbseries}, which is well-motivated by the physics
involved in the imaginary potential and also by perturbation theory,
it is possible to describe the Wilson loop data by other structures,
leading to different $\vim$. In particular, a very good description of
the data is provided by the form
\beq
\wrt \ = \ e^{- \vre \left(\tau \, - \, \frac{\beta}{2} \right) \; - \;
  \frac{\vim}{\pi} \, \int_{\beta/2}^\tau \, \log \,
  \frac{\beta-\tau}{\tau} - ....} \
W_{\scriptscriptstyle T}(\beta/2, \vec{r}) .
\eeq{logform}
The spectral peak obtained from this form is considerably different
from that shown above; see \fgn{logcot}. A Bayesian analysis, in our
opinion, ought to include the broad features of the low $\om$ peak
discussed in the previous paragraph. 

\bef
\centerline{\includegraphics[width=7.5cm]{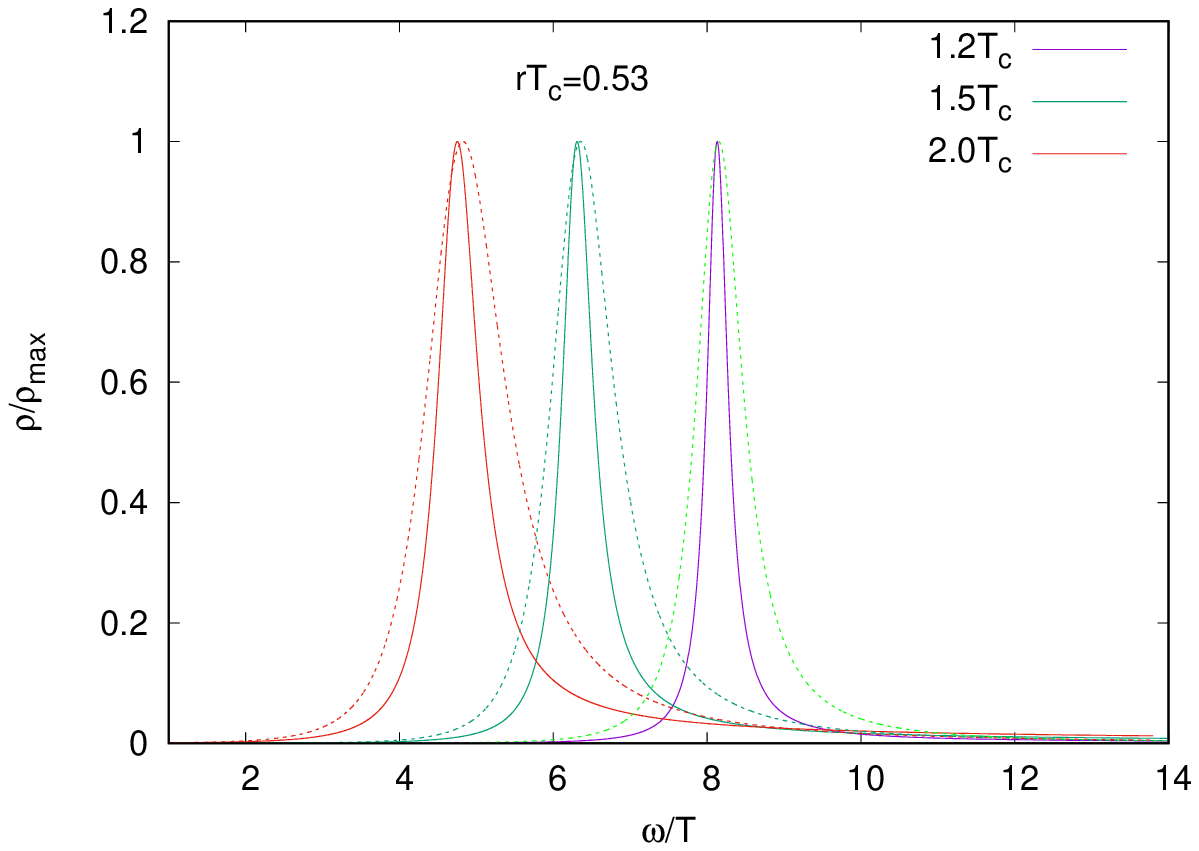}}
\caption{The low $\omega$ structure indicated by 
  \eqn{logform} (dotted line) compared with that obtained
  from \eqn{combo}, at $R$=8, at three different temperatures.}
\eef{logcot}

\section{Discussion of potentials and quarkonia}
\label{sec.pheno}
Let us try to analyze in some detail the potentials obtained in
\scn{results}. We start with $\vre$. \fgn{pot-real} shows our
estimation of $\vre$ at different temperatures. As is expected for a
gluonic plasma, the thermal effects are negligible at temperatures of
0.75 $\tc$ : the potential agrees completely between 0.75 $\tc$ and
0.63 $\tc$. So the potential at our lowest temperature measured for
each set can safely be
taken to approximate the zero-temperature potential.  The potential
shows the familiar features of the $1/r$ singularity at short
distances and the linear rise at large distances, and gives a good fit
to the Cornell form.

As we cross $\tc$, the finite temperature potential is close to that
at $T$=0 at short distances. But clear temperature effects are seen as
$r$ increases: in particular, the linear behavior of the $T=0$
potential gets screened.  In perturbation theory one expects, in
leading order, a Debye-screened form of $\vre$ which is same as the
free energy \cite{impot},
\beq
\vrp \ = \ - \frac{\alpha(T)}{r} e^{- \md r} - \md \alpha(T) + C
\eeq{vrp}
where $\md = g T$ in leading order and $\alpha(T)$ is the running
coupling at the appropriate temperature scale.
In \fgn{remod} this form,
\eqn{vrp}, is shown at different temperatures, along with the
nonperturbatively obtained potential. For drawing the perturbative
curve, following \cite{impot}, we have
used one-loop formula for the coupling \cite{kajantie},
$\alpha^{-1}(T) \ = \ \frac{\textstyle 33}{\textstyle 8 \pi} \,
    \log(6.742 \, T/\lms)$, and   $T_c/\Lambda_{\overline \rm MS}$ =
    1.10-1.20 \cite{largeN}. The band in the
perturbative form in \fgn{remod} corresponds to this range
in $T_c/\Lambda_{\overline \rm MS}$. Since we are interested in the $r$
dependence of $\vre$, the additive renormalization constant $C$ is
fixed by matching to the lattice potential at $r \tc$ = 0.5 at $T=2 \tc$.

As \fgn{remod} shows, the perturbative form does not explain the
potential obtained in \scn{real}. In particular, the long distance
part of the potential is not as flat as the screened Debye form
predicts: as if a shadow of the string tension rise survives.

\bef
\centerline{\includegraphics[width=6cm]{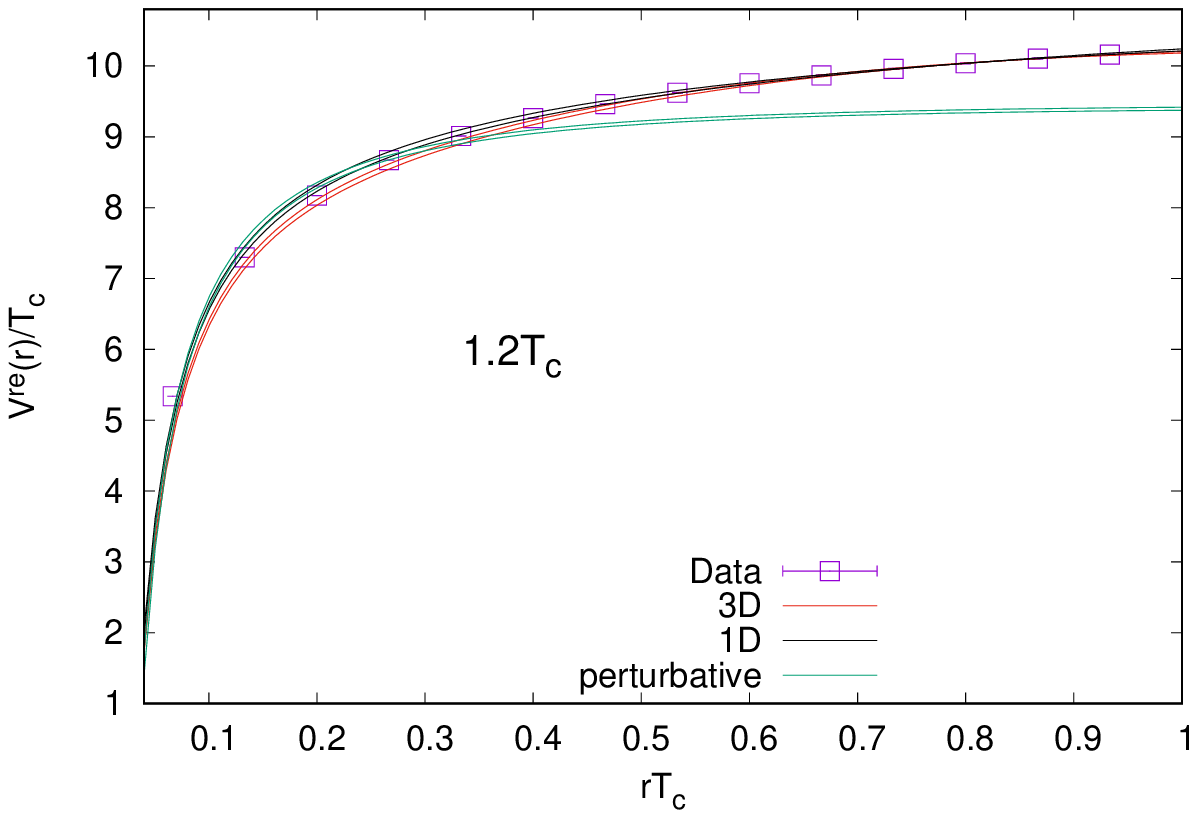}\includegraphics[width=6cm]{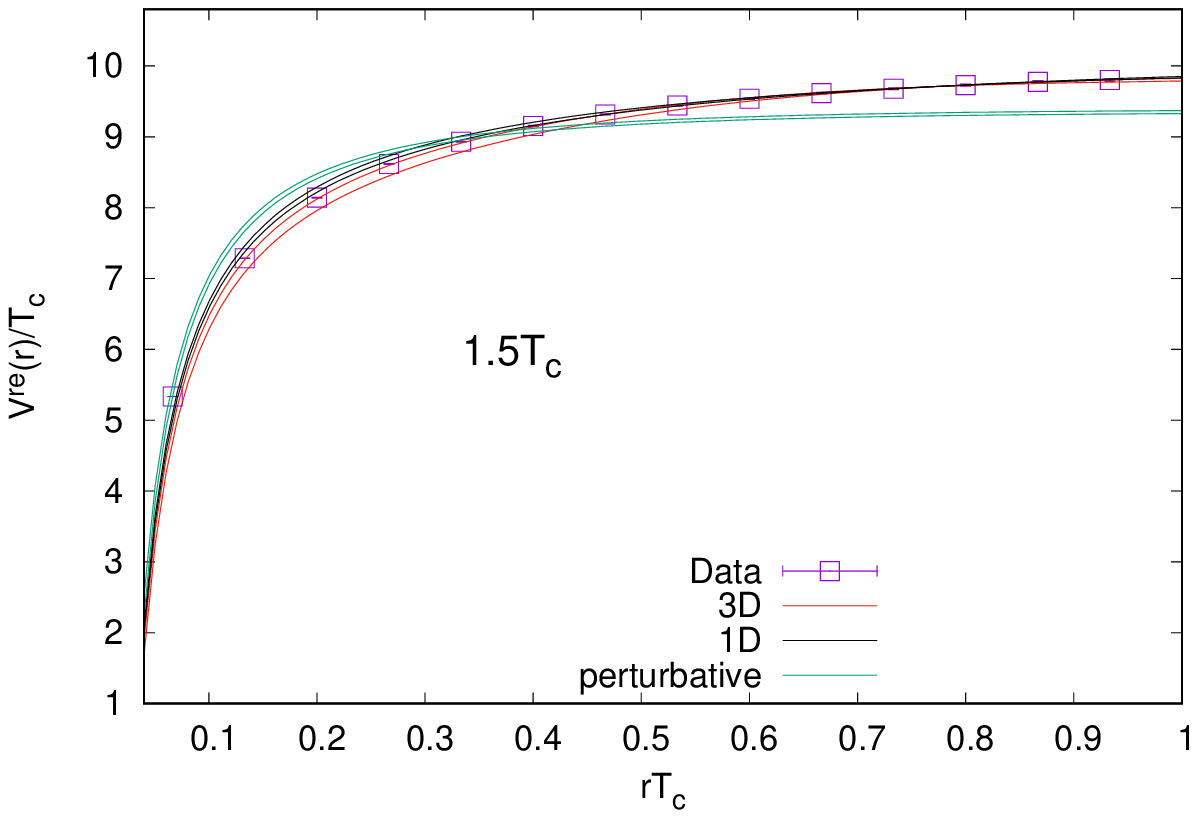}\includegraphics[width=6cm]{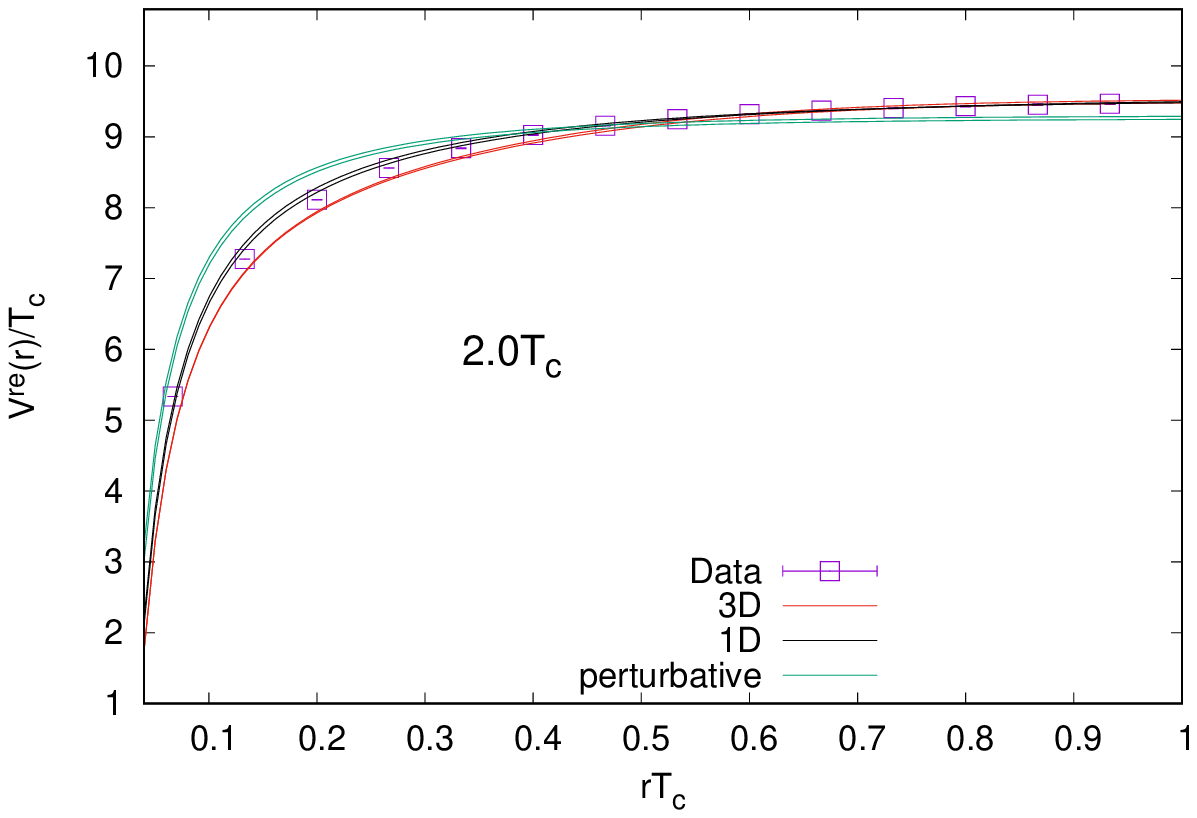}}
\caption{The finite temperature potential $\vre$ at temperatures 1.2
  $\tc$ (left), 1.5 $\tc$ (middle) and 2 $\tc$ (right), shown with
  various models for the potential: perturbative form (\eqn{vrp}),
  Debye-screened string in 1-d (\eqn{vro}) and in 3-d (\eqn{vrt}).} 
\eef{remod}

Since the long distance part of the $\qqb$ potential in QCD
vacuum has a linear string tension term, a natural next step would be
to try a screened form of the string tension term. The string tension
term being entirely nonperturbative, there is, however, no single
unique/preferred possibility for the screened form of this term. We
will consider here two models for screening that have been discussed in
the literature. A linear string tension is obtained in the 1+1
dimensional Schwinger model. Since string is essentially an
one-dimensional object, one can assume that the physics of screening
of the string term will also be similar to that in the Schwinger model. 
Such a consideration leads to the potential \cite{kms}
\beq
\vro \ = \ - \frac{\alpha}{r} \; e^{-\md r} \ + \ \frac{\sigma}{\md} \; \left(1
\; - \; e^{- \md r} \right) \ + \ C^\prime .
\eeq{vro}
This form of the screened potential can also be obtained by
generalizing the timelike gluon propagator to \cite{guo}
\beq
D(p_0=0, \vec{p}) \ \equiv \ \frac{1}{p^2 \, + \, \md^2} \ + \ \frac{2
  \sigma / \alpha}{(p^2 \, + \, \md^2)^2} .
\eeq{guo}
The second term gives a linear string term in the limit $\md \to 0$.
We treat \eqn{guo} here as a purely phenomenological construct to
model the screening in string tension term. 
In perturbation theory, one expects $\alpha$ to  be a function of $r$
and $T$. In the Cornell potential, however, one usually treats
$\alpha$ as a constant. We follow \cite{kms} and keep $\alpha, \sigma$
fixed to their $T=0$ value, the temperature dependence entering in
\eqn{vro} only through $\md$. The long distance part of the
potential $\vre$, \fgn{pot-real}, is fitted to \eqn{vro} to obtain
$\md, \; C^\prime$. $\vro$ does a good job of explaining the measured
potential as shown in \fgn{remod}. We have tried a few fit ranges covering
the large distance side of our measured potential. The band in \fgn{remod}
shows the variation of the fit parameters on shifting the fit range. The 
narrowness of the band is evidence for the stability of the fit to the form
of \eqn{vro}. The fitted value of $\md$ obtained from the fits is shown in 
\tbn{params}; the range corresponds to this change in fit range.

\bet
\setlength{\tabcolsep}{10pt}
\begin{tabular}{crllll}
   \hline
   & \multicolumn{2}{c}{$\md/T$} & & & \\ 
   $T/\tc$ & 1 D & 3 D & $b$ & $a_1$ & $a_2$ \\
   \hline
   1.2 & 1.18(6) & 1.37(6) & 0.34(1) & -1.23(7) & 1.22(4) \\
   1.5 & 1.34(8) & 1.49(6) & 0.35(3) & -0.87(2) & 1.85(2) \\
   2.0 & 1.46(8) & 1.60(9) & 0.29(2) & -0.8(2) & 2.85(38) \\
   \hline
\end{tabular} 
\caption{Various parameter sets related to the potential models
  discussed in \scn{pheno}. 1D refers to \eqn{vro}, 3D to \eqn{vrt}.
  $b, \, a_a, \, a_2$ are defined in \eqn{fitpot}.The errors shown include
the variation with fit range, and should be treated as a systematic
band rather than a statistical $1-\sigma$ band.
} 
\eet{params}

A different line of argument to a screened potential is to start with a
generalized Gauss' law which gives a linear potential \cite{dixit}. The
medium effect then can be incorporated by introducing a medium
permittivity \cite{patra}. Using an isotropic permittivity motivated
by HTL perturbation theory leads to the potential \cite{brpot}
\beq
\vrt \ = \ - \frac{\alpha}{r} \; e^{-\md r} \ - \ \frac{\Gamma(1/4)}{2
  \pi} \ \frac{\sigma}{\mu} \; \sqrt{x} \; K_{\frac{1}{4}}
\left(\frac{x^2}{2} \right) \ + \ C^{\prime \prime}
\eeq{vrt}
where $\mu^2 = \md \sqrt{\frac{\textstyle \sigma}{\textstyle \alpha}}$,
$x = \mu r$ and $K_{1/4}$ is the modified Bessel
function of the second kind \cite{fnote1}. At
large $r$, the second term behaves like $\frac{\textstyle
  \exp(-x^2/2)}{\textstyle \sqrt{x}}$. The results of the fit to this
form are also shown in \fgn{remod} and the value of $\md$ shown in
\tbn{params}. The fit to \eqn{vro} is found to be slightly more stable
than that to \eqn{vrt}, and so we use it for analysis of 
quarkonia behavior. However, \eqn{vrt} also approximately captures the
$r$ dependence of $\vre$; with our data we can not
statistically rule out either of the one-dimensional and
three-dimensional screening forms.

The imaginary part, $\vim$, turns out to be more difficult to model
using the conventional screening forms available in the literature. 
The perturbative form of the imaginary part, \eqn{htl}, is shown
in \fgn{immod} together with our data, for three different
temperatures. The parameters used are identical to that for the real
part, as detailed below \eqn{vrp}. The data shows very different
behavior from that of \eqn{htl}: at short distance, the perturbative
result overshoots the data, but it soon saturates, while our
nonperturbative data does not show a sign of saturation in the
distance scale studied by us. The perturbative result $ \vim_{\rm
  pert}$ behaves $\sim r^2 \log r$ at small $r$, and saturates to
$\sim \alpha \ T$ as $r \to \infty$. The nonperturbative
data shows a $r^2$ behavior to a much larger distance: in
particular, almost the whole range of $r$ explored by us,
$r \tc \lesssim 1$, can be fitted to a quadratic behavior at 1.5 $\tc$
and 2 $\tc$. 

The HTL permittivity that leads to \eqn{htl}, is complex, so as to
produce a complex potential. Use of this permittivity in the
generalized Gauss' law leads to \cite{brpot} \eqn{vrt} and an
imaginary part 
\bea
\vit &=&  \vim_{\rm pert} \ + \ \alpha \, T \, \left\{ D_{-1/2}
\left(\sqrt{2} x \right)
\int_0^x dy \; {\rm Re} \, D_{-1/2} \left(i \sqrt{2} y \right) \, y^2 \,
g\left(\frac{\md}{\mu} \, y \right) \right. \label{vit} \\
&+& \left. {\rm Re} D_{-1/2} \left(i \sqrt{2} x \right) \int_x^\infty dy \,
    D_{-1/2} \left(\sqrt{2} y \right) y^2 \, g \left(\frac{\md}{\mu} \, y \right)
    \ - \ D_{-1/2}(0) \, \int_0^\infty dy \; D_{-1/2}\left(\sqrt{2} y \right)
    \, y^2 \, g\left(\frac{\md}{\mu} \, y \right) \right\} \nonumber \\
    {\rm where} \  g(x) & = & \int_0^\infty dz \; \frac{\textstyle 2 \, z}
    {\textstyle z^2 \, + \, 1} \ \frac{\textstyle \sin z x}{\textstyle
      z x}. \nonumber
\eea
$\vit$ is also shown in \fgn{immod}, with the
legend `3D'. Here the value of $\md$ obtained
from \eqn{vrt} is used, and the band corresponds to the range in
$\md$ (\tbn{params}). This form has a similar behaviour $\sim r^2$ at
small $r$ to the data. While it is steeper at large $r$ than the
perturbative form, it is less steep than our data.  

If one uses a complex permittivity analogous to the HTL term
in conjunction with the modified propagator of \eqn{guo}, one can get
the ``complex potential'' for 1D screening, i.e., the imaginary  part
of \eqn{vro}. The imaginary part reads \cite{guo}  
\beq
\vio \ = \ \vip \ + \vis, \qquad
\vis \ = \ \frac{4 \sigma T}{\md^2} \ \int_0^\infty dz \,
\frac{2 z}{(z^2+1)^3} \left[ 1 - \frac{\sin z x}{z x} \right] .
\eeq{vio}
$\vio$ is shown in \fgn{immod} with legend `1D'; the value of $\md$
is that obtained from \eqn{vro} in \tbn{params}. This form seems to
have a higher slope than our data at small $r$ and a smaller slope at
large $r$, though at 1.5 $\tc$ it is close to our data in the range
of $r$ studied by us.

\bef
\centerline{\includegraphics[width=6cm]{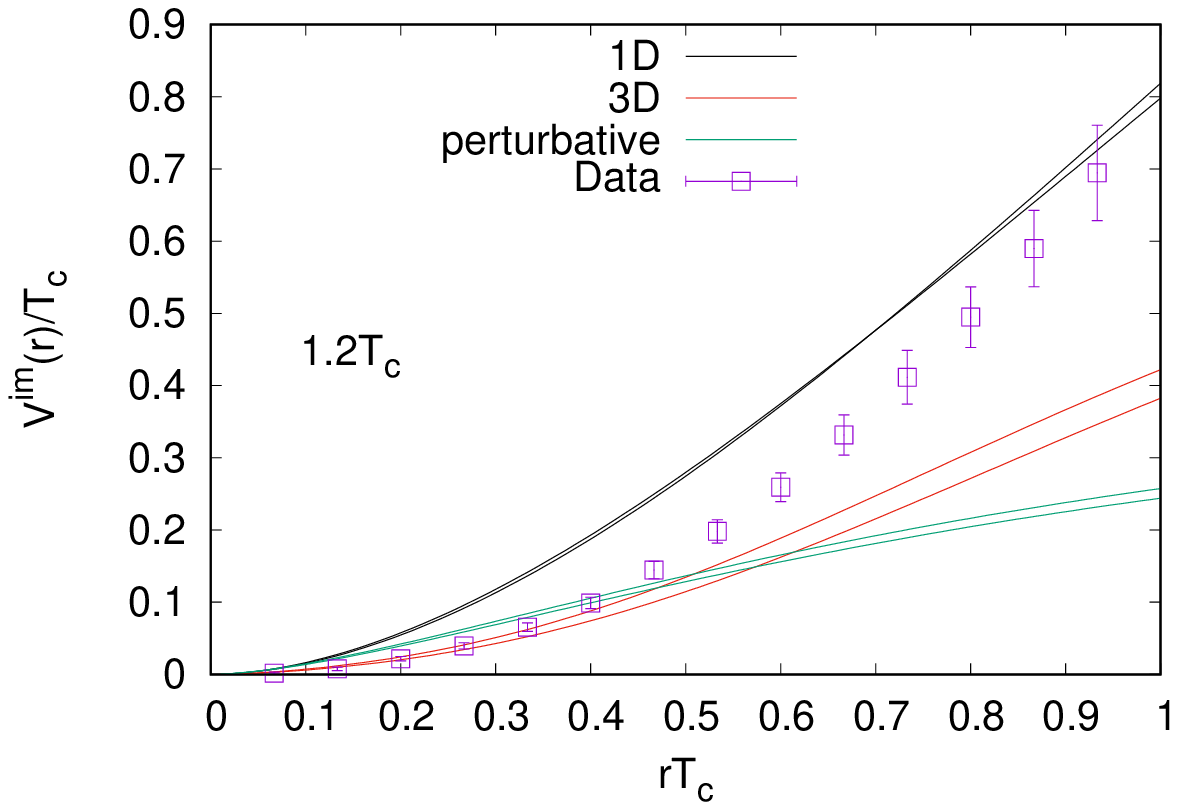}
  \includegraphics[width=6cm]{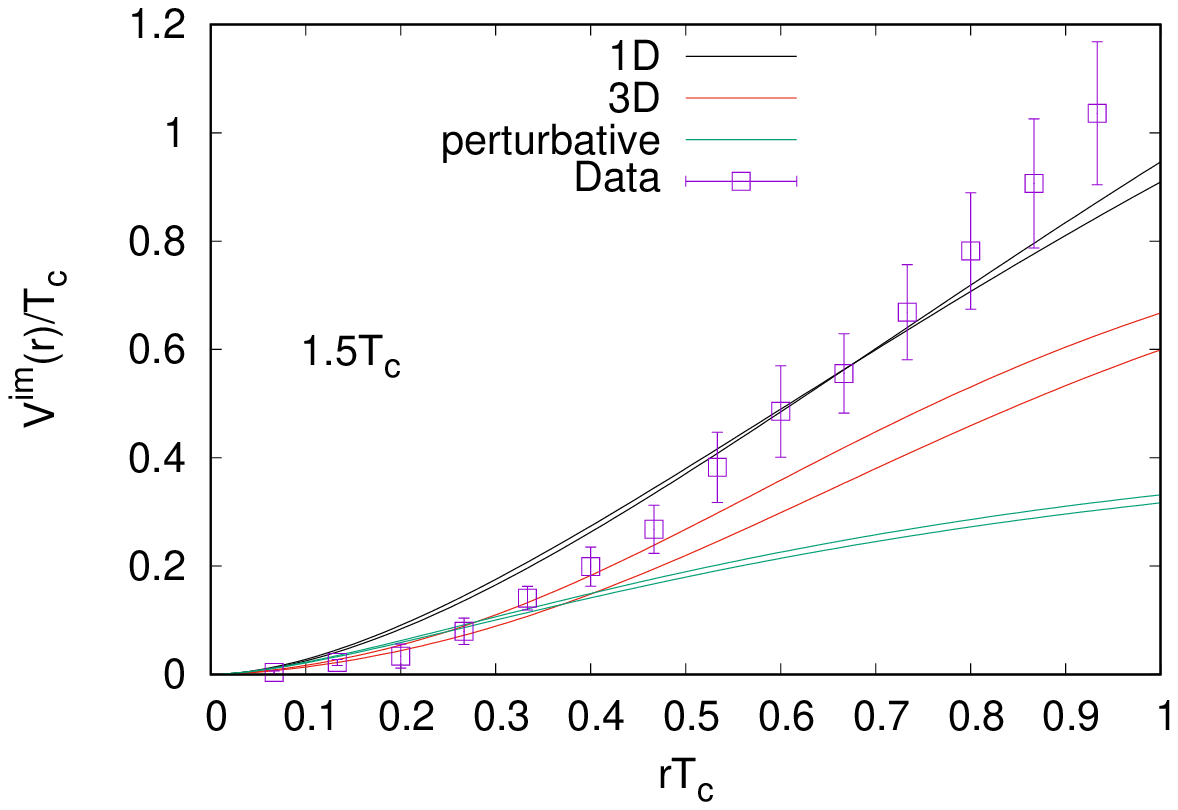}
  \includegraphics[width=6cm]{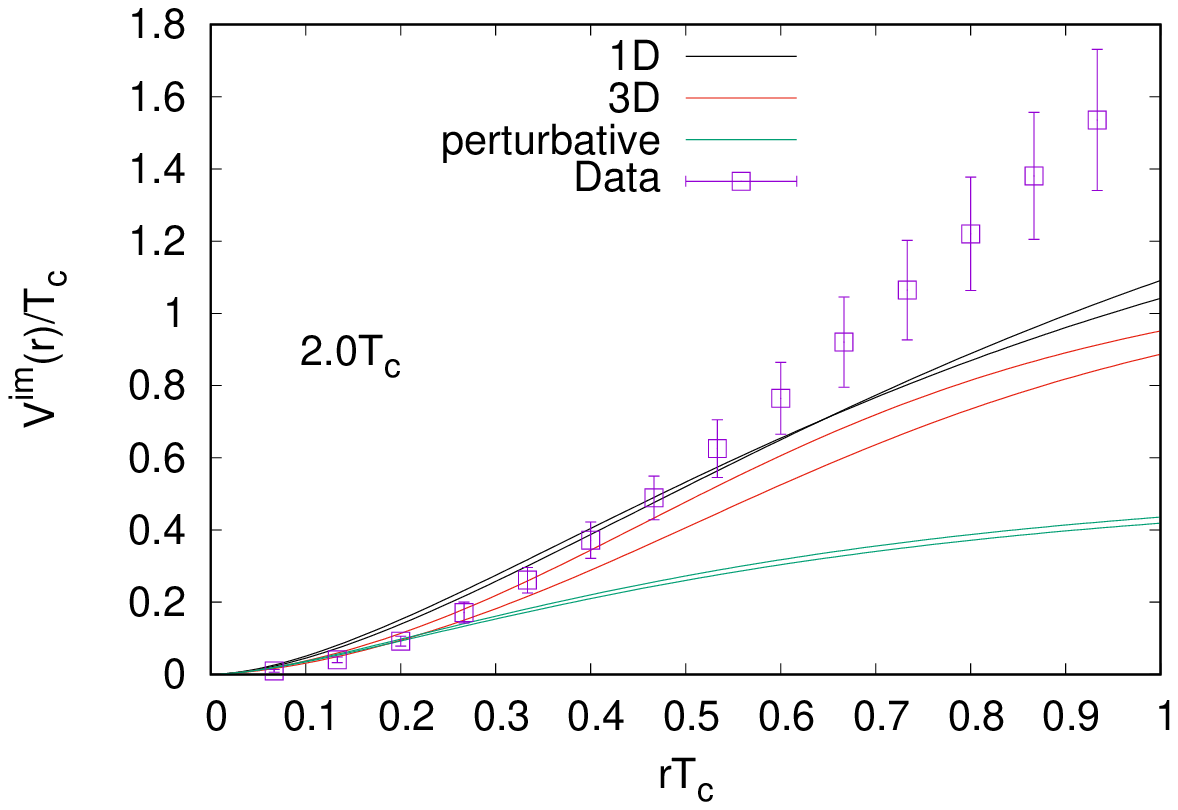}}
\caption{The imaginary part of the finite temperature potential,
  $\vim$, at temperatures 1.2
  $\tc$ (left), 1.5 $\tc$ (middle) and 2 $\tc$ (right), shown with
  various models for the potential: perturbative form (\eqn{htl}),
  Debye-screened string in 1-d (\eqn{vio}) and in 3-d (\eqn{vit}).}
\eef{immod}

As \fgn{immod} reveals, none of the simple forms discussed does a good job
of modelling our data for the imaginary potential over the range of
$r$ studied by us. At small $r$, the numerically calculated potential
has a smaller slope than either the screened string forms or the forms
\eqn{vio} and \eqn{vit}. At large $r$, on the other hand, it is
steeper. Both of these latter forms, in turn, show a
much larger imaginary part than the perturbative form at large $r$,
with \eqn{vit} comparable to our data at larger values of $r$.

We are interested in the ground state quarkonium peaks in the
spectral function. While it is most sensitive to the short distance
part of the potential, it is also affected by the long distance part,
especially as the binding energy becomes less and the state becomes
broader. As we mentioned before, in the range $r \tc \lesssim 1$
studied here, our data for $\vim$ grows $\sim r^2$. Of course, on
physical principles we expect it to saturate at large $r$. Motivated
by \eqn{vio}, we tried to model the imaginary part of the potential by
fitting the data to an arbitrary combination of $\vip$ and $\vis$.
We also tried to fit it to a purely quadratic form. Finally, 
we calculate the spectral function for finite mass quark
through integrating \eqn{nr}, with
\beq
V_{\qqb}(\vec{r}, T) \ = \ \vro \ - i \ \vif, \qquad \vif \ =
\ \begin{cases} b \, T \, (r T)^2 \\ a_1 \, \vip \; + \; a_2 \,
\vis \end{cases}
\eeq{fitpot}
where $\vro$ is given in \eqn{vro} and the parameters $b, a_1, a_2$
are given in \tbn{params}. We emphasize that our forms for $\vif$
represent purely phenomenological fits of the data; one can take them
to correspond to two limiting asymptotic behaviors given the
data. We will treat the results for the spectral function obtained
with the two forms of $\vim$ in \eqn{fitpot} as a systematic band, and
look for features of the band.

In the left panel of \fgn{rhoM} we have shown
the spectral function obtained this way at 1.5 $\tc$, with the quark
mass varying from 1.5 GeV to 6 GeV. At $T=0$, using the unscreened
Cornell potential we get a series of sharp peaks. We denote the mass
of the 1S state as $M_P$, and normalize the $x$ axis with respect to it
in \fgn{rhoM}. At 1.5 $\tc$, even
for a quark mass of 6 GeV we only find one peak. Of course, 1.5 $\tc$
here corresponds to a temperature of about 420 MeV. Expectedly, the
peak is the sharpest for the heaviest quark, gradually
broadening till, for quark masses close to the charm, only a very
broad peak structure can be seen. The spectral function for $\mq$ =
1.5 GeV is also qualitatively different from the others, and is
very different from the spectral function obtained directly from
$J/\psi$ correlators in \cite{mem}, but in qualitative agreement with
a later study \cite{hengtong}. Similar results have been
seen in \cite{mocsy}. In the right panel of
\fgn{rhoM} we have shown the results for the $\Upsilon$ peak.
Quark mass $\mq$ was tuned to get the 1S meson mass $\sim$ 9.45 GeV.
A sharp peak is seen at 1.2 $\tc$, which gradually broadens as the
temperature increases. But
a peak structure survives all the way to 2 $\tc$. Note that 2 $\tc$
here corresponds to about 560 MeV, setting the scale using the string
tension. Also at low temperatures the peak is quite narrow, in
comparison to what was found from nonrelativistic bottomonia
correlators in \cite{memnr1}.  

\bef
\centerline{\includegraphics[width=7.5cm]{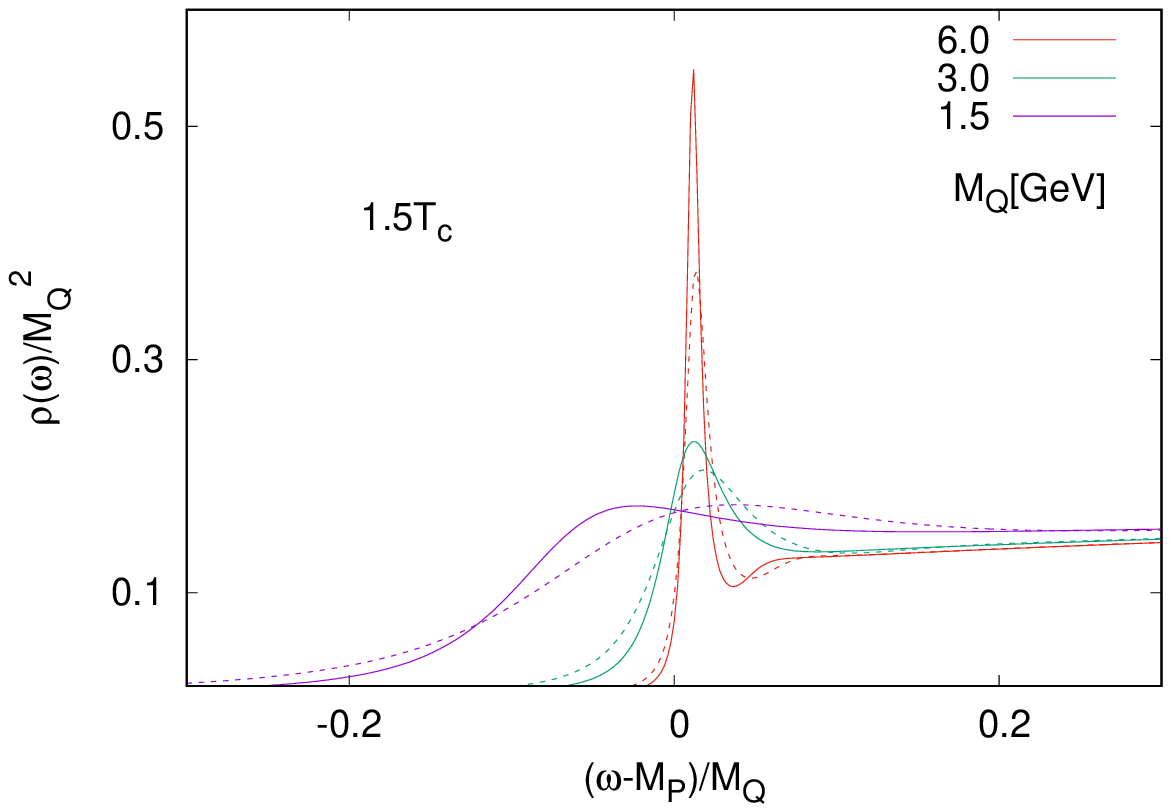}
  \includegraphics[width=7.5cm]{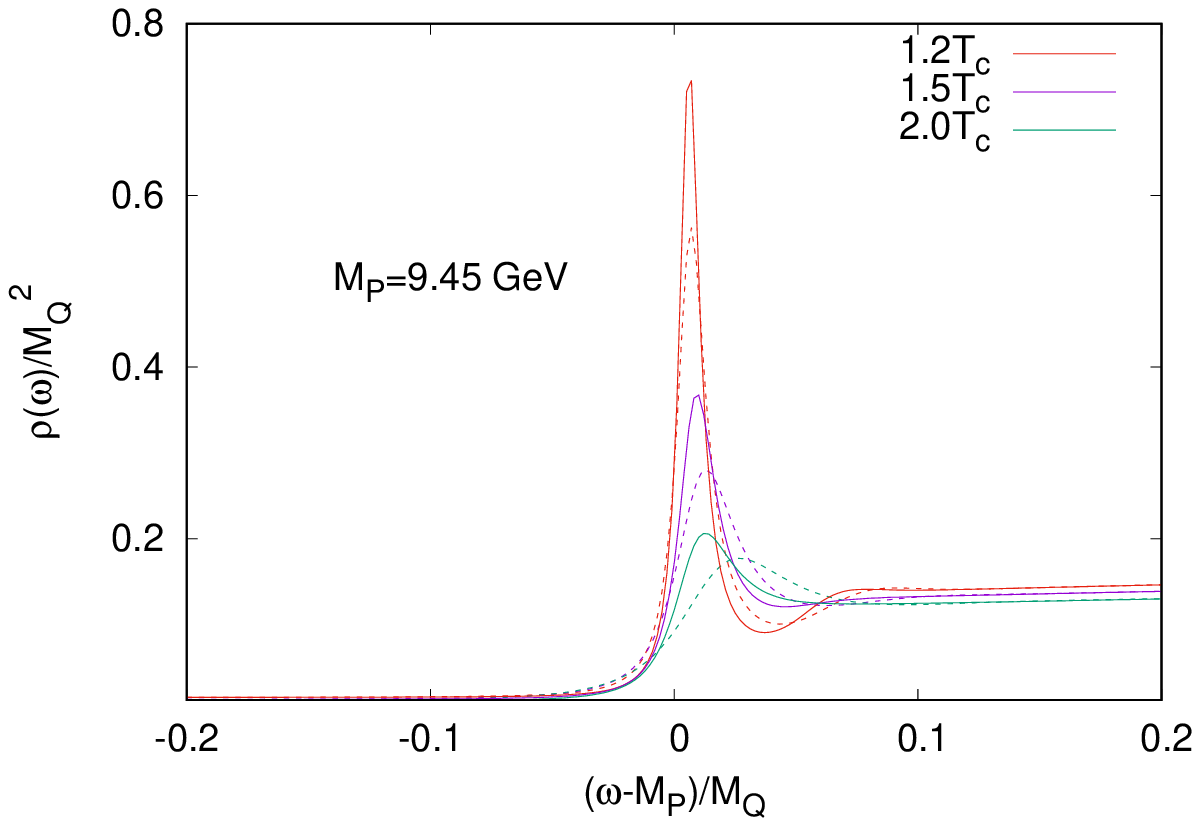}}
\caption{(Left) the peak of the spectral function $\rho(\omega,
  \vec{r} \to 0)$ at a temperature of 1.5 $\tc$, for different values
  of the quark mass. The dashed line coresponds to the quadratic form
  and the full line, to the two-parameter form of $\vim$ in
  \eqn{fitpot}. (Right) The spectral function at different
  temperatures for a quark mass close to the bottom, M(1S)=9.45 GeV.
  In the $x$ axis, the zero is at $\om=M_P$, the 1S state mass
  obtained using the Cornell potential.}
\eef{rhoM}

\section{Summary}
\label{sec.summary}
One popular way of studying the medium modification of quarkonia in
quark-gluon plasma is through defining an effective ``in-medium''
potential.  Theoretically, a suitable potential can be defined
\cite{impot,blaizot7} by examining the time dependence of the
Minkowski-time Wilson loop, \eqn{pot}. This potential is complex, with
the real part of the potential describing the Debye-screened binding
of the $\qqb$ pair in medium and the imaginary part related to damping
of the wavefunction due to interaction with the thermal medium. A
nonperturbative extraction of this potential \cite{rhs} involves
extracting the low-frequency structure of the spectral function from
the Euclidean-time Wilson loop, \eqn{poteucl}. This is in general a
very difficult problem. The existing studies in the literature have
either progressed through using Bayesian analysis, with their
associated, and sometimes hard-to-estimate, systematic errors, or by
making ad-hoc assumption about the low-frequency structure.

In this work we have introduced a new method of nonperturbative evaluation
of the potential. We find that a reorganization of the Euclidean Wilson
loop data, motivated by the underlying structure of the finite temperature
correlation function (see \apx{pert}),
leads to an enormous simplification in the extraction of the
peak structure from the Wilson loop. The main ingredients of our
method are outlined in \scn{def}, and the details are given in
\scn{results}. We have calculated the potential in a gluonic plasma
for temperatures $\le 2 \tc$ from smeared Wilson loops, calculated
using anisotropic lattice discretization of
the gluonic theory. Our results for the potential are summarized in
\fgn{pot-real} and \fgn{pot-im}. The real part of the potential, which
shows the Cornell form below $\tc$ with no noticable temperature
dependence upto 0.75 $\tc$, shows Debye screening on crossing $\tc$,
with the screening mass increasing with temperature. The form of the
potential is different from the perturbative form at least upto 2
$\tc$, as illustrated in \fgn{remod}. The imaginary part of the
potential is zero below $\tc$. Above $\tc$ it rises rapidly, with a
spatial dependence $\sim r^2$ till distances $r \lesssim 1/\tc$. Its
behaviour is sharply different from the perturbative result, as
illustrated in \fgn{immod}.

In the course of our study, we have also investigated issues like
dependence of the finite-temperature potential on the definition of
the operator, which, we feel, have not been properly discussed in the
literature. We have examined how the potential depends on the
smearing, and compared the potential obtained from smeared Wilson
loops with those from Coulomb gauge fixed Wilson line correlators. We
have also examined the relation between the real part of the potential
and the free energy of a static $\qqb$ pair in the plasma; see
\fgn{free-Vre}. In \scn{combo} we have discussed the structure of the
low energy peak of the spectral function. It is quite different from
the Lorentzian structure that has often been assumed in direct
extractions of potential from Euclidean wilson loop using
\eqn{poteucl}. We have also
illustrated, with example, the difficulty of extracting the low energy
peak from the Euclidean Wilson loop without putting in additional
physics input.

Our data for the extracted potential can be found in
\scn{results}, in particular in \fgn{pot-real} and \fgn{pot-im}.
Moreover, for various purposes it is convenient to have a
parametrization of the potential. In \scn{pheno} we have explored
various standard forms of a screened potential. As \fgn{remod} shows,
for the real part, the
form of 1D screening of the string potential seems to give a
reasonable description of the data, with parameters given in
\tbn{params}. For the imaginary part it is more difficult to find
quantitative agreement with a standard screened form. The potential
rises $\sim r^2$ till intermediate distances $r \tc \sim 1$. While the
potential is expected to saturate as $r \to \infty$, it is difficult
to make any statement about that behavior from our data at $r \tc
\lesssim 1$.  A purely phenomenological generalization of \eqn{vio}, using
arbitrary linear combination of $\vim_{\rm pert}$ and $\vim_\sigma$, seems to
give a good description of the data in the range of $r$ explored by
us, with coefficients given in \tbn{params}. 
Since we expect the long distance behavior of $\vim$ to be somewhere
between this and the $r^2$ behavior, for a study of quarkonia in
the plasma we use both of the forms for $\vim$ (see \eqn{fitpot}). The
difference in the spectral structure obtained with these two forms is
considered as a qualitative systematic band.

The spectral function peaks for S-state quarkonia with different quark masses
are shown in \fgn{rhoM}. In the left panel, the variation of the
spectral function with $\mq$ is shown. Below $\tc$
the spectral function has a number of narrow peaks corresponding to
the nS states; but above $\tc$ only the 1S peak survived even for $\mq$ =
6 GeV. For $\mq$ = 1.5 GeV, close to the mass of charm, there is no
significant peak structure at this temperature. Of course, the
nonrelativistic formalism may not be valid for charmonia at these
temperatures. For $\mq$ = 3 GeV, a clear peak structure is seen at 1.5
$\tc$, with very little shift in the peak position. In the right panel
of \fgn{rhoM} the spectral function for 1S bottomonia is shown. While
the peak structure weakens with temperature, a clear peak survives
till 2 $\tc$, with very little shift in peak position, and reasonably
narrow peak, at least till 1.5 $\tc$.

While potential by itself does not provide a complete description of
medium interaction of quarkonia, it is an important part of a complete
description, and can provide useful inputs for more sophisticated
studies like direct extraction of spectral functions; they also can
provide essential nonperturbative ingredients of an open quantum
system analysis of in-medium quarkonia \cite{blaizot15, miura}.  Our
results are for the quenched theory, and one needs to be careful when
applying them for quarkonia phenomenology. They, however, provide
benchmarks for comparing with direct extractions of quarkonia spectral
functions from euclidean correlators. More importantly, the method we
have outlined for the extraction of the potential, \scn{def}, is quite
simple and stable, and we expect one should be able to use it to
extract reliable potential also from dynamical lattices.

{\bf Acknowledgements:} This work was carried out under the umbrella
of ILGTI.  The computations reported here were performed on the 
clusters of the Department of Theoretical Physics, TIFR. We would like
to thank Ajay Salve and Kapil Ghadiali for technical support. DB would like
to thank Rajiv Gavai, Alexander Rothkopf and Peter Petreczky for discussions.

\appendix
\section{Lattice parameters}
\label{sec.sets}
We use the anisotropic Wilson gauge action for our discretization of the 
gluonic theory. The discretized euclidean action is
\beq
S_G \ = \ \frac{\bs}{N} \sum_{ij} {\rm Re} \; \tr \left( 1 - P_{ij} (x) \right) 
+ \frac{\bt}{N} \sum_i {\rm Re} \; \tr \left( 1 - P_{4i} (x) \right) 
\eeq{aniso}
where $i,j$=1,2,3 and $P_{\mu \nu}$ are the plaquette variables in the 
$\mu, \nu$ direction. Since we are interested in fine grating in the time
direction, we use $\bt \gg \bs$. 

Klassen has provided a convenient way of non-perturbatively finding 
couplings suitable for an anisotropy, $\xi=\as/\at$ from comparison 
of spatial and space-time Wilson loops \cite{klassen}: find 
$\xi_0=\sqrt{\bt/\bs}$ such that
\beq
\frac{R(x,y)}{R(x,t=\xi y)} \; = \; 1  
\qquad {\rm where} R(i,j) \ = \ \frac{W(i,j)}{W(i+1,j)}.
\eeq{klasaniso}
An interpolating formula for
estimating $\beta_s$ and $\beta_\tau$ for a given $\xi$ is also given
\cite{klassen}. We use this formula to get the suitable couplings for
our purpose and then checked the anisotropy by comparing the
potentials as mentioned above.We use this formula to get a set of prior
guesses for lattices with $\xi$=3, and then did some simulations to
tune the anisotropy. The final parameter set for our runs is shown in
\tbn{sets}.

For each of the lattices in \tbn{sets} we have used about $\mathcal{O}(10^4)$
configurations. For each set, the configurations were generated from 90
independent (different random number seeds) runs. For the runs below
$\tc$, multilevel algorithm was used. About 2000 measurements were
collected (90 independent runs with 20-25 measurements each), with
each measurement being sublattice average of 200-400 updates, and
after each measurement 100 decorrelating sweeps were made. Each sweep
consisted of 1 heatbath and 3 overrelaxation steps (this was kept
fixed across all runs). Above $\tc$ multilevel was less
cost-effective, and was used mostly for the ruls with large $\nt$. The
multilevel steps were similar to what is described above, but a larger
number of measurements $\sim 4500-9000$ were used. In the runs above
$\tc$ where multilevel was not used, 9000-27000 configurations were
used, two configurations being separated by 100 sweeps.

\section{Perturbative expressions}
\label{sec.pert}
In \scn{def} we have outlined our method to extract the thermal
potential, \eqn{nr}. At $T=0$ the definition of $V(\vec{r})$ through
\eqn{pot} is well-understood
diagrammatically: the ladder of time-ordered gluon
propagators $D_{00}$ (including the crossed diagrams) lead to an
exponentiation of the Fourier transform of $D_{00}(0, \vec{k})$, which
defines the potential \cite{fischler}.

At finite temperature, the structure of the Wilson loop is more
complicated. It was stressed in \cite{blaizot7}, however (and
demonstrated for QED) that in order for a potential to exist
via \eqn{pot} the ladder of the time-ordered gluon propagators
need to be resummed. Then $W_M(t, \vec{r})$ will have the structure
\beq
W_M(t, \vec{r_1}, \vec{r_2}) \ \sim \ e^{-i \, \int_0^t \, dt_1 \;
  \int_0^t \, dt_2 \; D^{00}_T(t_1-t_2, \, \vec{r_1} - \vec{r_2} ) }.
\eeq{wmpot}
Here we have only shown the potential part that depends on $\vec{r_1}
- \vec{r_2}$, omitting self-energy corrections and non-potential
contributions.  

The Euclidean wilson loop can, similarly, be written as
\cite{blaizot7} 
\beq
\wrt \ \sim \ e^{- \int_0^\tau d\tau_1 \; \int_0^\tau d\tau_2
  \Delta(\tau_1-\tau_2, \vec{r})}
\eeq{wpot}
where the finite-temperature imaginary-time propagator has the
structure \cite{blaizot7}
\beq
\Delta(\tau, \vec{r}) \ = \ \int \frac{d \omega}{2 \pi} \; e^{-\om \tau}
\; \rho_{\scriptstyle D}(\om, \vec{r}) \; \left[ \theta(\tau) + \nbw \right].
\eeq{prop}
Here we have used a mixed representation in the right hand side:
$\rho_{\scriptstyle D}(\om, \vec{r})$ is the spatial Fourier transform of the usual spectral
function.

Putting \eqn{prop} in \eqn{wpot} gives, using $\nbw = e^{\textstyle
  -\beta \om} (1+\nbw)$, 
\beq
\log \wrt \ \sim \ \int \frac{d\om}{2 \pi} \; \tau \; 
\frac{\rho_{\scriptstyle D}(\om, \vec{r})}{\om} \ + \ \int \frac{d\om}{2 \pi} \,
\left(1+\nbw \right) \left( e^{-\omega \tau} \, + \, e^{-\omega (\beta
  - \tau)} \right)  \frac{\rho_{\scriptstyle D}(\om,
  \vec{r})}{\om^2},
\eeq{diag}
omitting $\tau$ independent terms. The first and second terms in the right
hand side of \eqn{diag} correspond to $\btau$ and $\atau$ in \scn{def}.

For QCD, the expression for the Wilson loop has been calculated in
\cite{impot} to leading order in HTL perturbation theory. For convenience,
we reproduce here the results of \cite{impot}, written in the notation of
\scn{def}.
\bea
\atau &=& 2 g^2 c_f \int {\bf dq} \ \sin^2 \frac{q_3 r}{2}
\ (1 \, + \, \nbw) \; \left(e^{- \omega
  \tau} \, + \, e^{-(\beta-\tau) \om} \right) \nonumber \\
      &\times& \left\{\left(\frac{1}{\vec{q}^2} \, - \,
\frac{1}{\om^2} \right) \rew \ + \ \left(\frac{1}{\vec{q_3}^2} \, -
\, \frac{1}{\vec{q}^2} \right) \rtw \right\} \ + \ \tau \ {\rm
  indep. \ terms} \label{pert} \\  
\btau &=& 2 g^2 c_f \int {\bf dq} \ \sin^2 \frac{q_3 r}{2}
\ \frac{\beta/2 \, - \, \tau}{\om} \ \rew . \nonumber
\eea
Here ${\bf dq} = \frac{\textstyle d^3q}{\textstyle (2 \pi)^3} \;
\frac{\textstyle d \om}{\textstyle \pi}$, 
$c_f=\frac{\textstyle 4}{\textstyle 3}$ is the color factor, and $\rtw$,
$\rew$ are the spectral functions corresponding to the transverse and the
longitudinal parts of the gluon propagator. 

We are interested in the energy regime $\lvert \omega \rvert \ll
\lvert \vec{q} \rvert$. In this regime, the spectral functions
$\rew, \rtw$ become, to leading order in HTL perturbation theory,
\bea
\rew &=& - \pi \md^2 \frac{\om}{2 \lvert \vec{q} \rvert \; \left(\om^2
  \, + \, \md^2 \right)^2}, \nonumber \\
  \rtw &=&   \pi \md^2 \frac{\om}{4 \lvert \vec{q} \rvert^5}.
  \label{pertspectral}
\eea

\end{document}

%% file: macros.tex
\newcommand\bef{\begin{figure}}
\newcommand\eef[1]{\label{fg:#1}\end{figure}}
\newcommand\beq{\begin{equation}}
\newcommand\eeq[1]{\label{#1}\end{equation}}
\newcommand\bea{\begin{eqnarray}}
\newcommand\eea{\end{eqnarray}}
\newcommand\bet{\begin{table}}
\newcommand\eet[1]{\label{tbl:#1}\end{table}}

\newcommand\fgn[1]{Figure \ref{fg:#1}}
\newcommand\eqn[1]{Eq.\ (\ref{#1})}
\newcommand\scn[1]{Sec. \ref{sec.#1}}
\newcommand\apx[1]{Appendix \ref{sec.#1}}
\newcommand\tbn[1]{Table \ref{tbl:#1}}

\newcommand\jhep{{\sl J.\ H.\ E.\ P.\/}\ }
\newcommand\np{{\sl Nucl.\ Phys.\/}\ }
\newcommand\npps{{\sl Nucl.\ Phys.\ Proc.\ Suppl.\/}\ }

\newcommand\pr{{\sl Phys.\ Rev.\/}\ }

\newcommand\plt{{\sl Phys.\ Lett.\/}\ }
\newcommand\prt{{\sl Phys.\ Rept.\/}\ }
\newcommand\rmph{{\sl Rev.\ Mod.\ Phys.\/}\ }
\newcommand\epj{{\sl Eur.\ Phys.\ J.}\ }
\newcommand\ijmp{{\sl Int.\ J.\ Mod.\ Phys.}\ }
\newcommand{\lqcd}{\Lambda_{\scriptscriptstyle \rm QCD}}
\newcommand{\lms}{\Lambda_{\overline \scriptscriptstyle \rm MS}}
\newcommand{\mq}{M_{\scriptscriptstyle Q}}
\newcommand{\qqb}{Q \bar{Q}}
\newcommand{\at}{a_\tau}
\newcommand{\as}{a_s}
\newcommand{\bs}{\beta_s}
\newcommand{\bt}{\beta_\tau}
\newcommand{\tc}{T_c}
\newcommand{\nt}{N_\tau}

\newcommand{\om}{\omega}

\newcommand{\cgt}{C_>(t, \vec{r})}
\newcommand{\vt}{V_{\scriptscriptstyle T}(\vec{r})}
\newcommand{\nbw}{n_{\scriptscriptstyle B}(\omega)}
\newcommand{\fre}{F(\vec{r}; T)}
\newcommand{\vre}{V^{\rm re}_{\scriptscriptstyle T}(\vec{r})}
\newcommand{\vim}{V^{\rm im}_{\scriptscriptstyle T}(\vec{r})}
\newcommand{\vrp}{V^{\rm re}_{\rm pert}(\vec{r}, T)}
\newcommand{\vro}{V^{\rm re}_{\rm 1D}(\vec{r}, T)}
\newcommand{\vrt}{V^{\rm re}_{\rm 3D}(\vec{r}, T)}
\newcommand{\vio}{V^{\rm im}_{\rm 1D}(\vec{r}, T)}
\newcommand{\vit}{V^{\rm im}_{\rm 3D}(\vec{r}, T)}
\newcommand{\vip}{V^{\rm im}_{\rm pert}(\vec{r}, T)}
\newcommand{\vis}{V^{\rm im}_\sigma(\vec{r}, T)}
\newcommand{\vif}{V^{\rm im}_{\rm fit}(\vec{r}, T)}
\newcommand{\md}{m_{\scriptscriptstyle D}}
\newcommand{\rw}{\rho(\omega; T)}
\newcommand{\sw}{\sigma(\omega; T)}
\newcommand{\rew}{\rho_{\scriptscriptstyle E} (\om, \vec{q})}
\newcommand{\rtw}{\rho_{\scriptscriptstyle T} (\om, \vec{q})}
\newcommand{\rwr}{\rho(\omega, \vec{r}; T)}
\newcommand{\rjwr}{\rho_{\scriptscriptstyle J}(\omega, \vec{r}; T)}

\newcommand{\wrt}{W_{\scriptscriptstyle T}(\tau, \vec{r})}
\newcommand{\wrb}{W_{\scriptscriptstyle T}(\beta-\tau, \vec{r})}
\newcommand{\wpr}{W^p_{\scriptscriptstyle T}(\tau, \vec{r})}
\newcommand{\wap}{W^a_{\scriptscriptstyle T}(\tau, \vec{r})}

\newcommand{\atau}{A(\tau, \vec{r})}
\newcommand{\btau}{B(\tau, \vec{r})}
\newcommand{\bpom}{\bar{\psi}_{\scriptscriptstyle \Omega}}
\newcommand{\pom}{\psi_{\scriptscriptstyle \Omega}}
\newcommand{\wil}{\mathbb{U}}

%% file: paper.bbl
\begin{thebibliography}{99}
\bibitem{fnote1}
  A somewhat similar form, with
  however $K_{1/4}(x^2)$, is obtained by introducing the medium effect
  through linearized Boltzmann distribution instead \cite{dixit}.
\end{thebibliography}
